\documentclass[12pt]{iopart}

\expandafter\let\csname equation*\endcsname\relax
\expandafter\let\csname endequation*\endcsname\relax
\usepackage{amsmath}

\usepackage{amsfonts}
\usepackage{amssymb}
\usepackage{amsthm}
\usepackage{color}
\usepackage{bbm,dsfont}
\usepackage{graphicx}
\usepackage{hyperref}

\usepackage{enumerate}

\theoremstyle{definition}

\newtheorem*{definition*}{Definition}

\newcommand{\R}{\mathbb{R}} %real
\newcommand{\half}{\tfrac{1}{2}} %half
 
\newcommand{\hi}{\mathcal{H}} %Hilbert space H
\newcommand{\hik}{\mathcal{K}} %Hilbert space K
\newcommand{\lh}{\mathcal{L(H)}} %bounded linear operators
\newcommand{\lk}{\mathcal{L(K)}} %bounded linear operators on K
\newcommand{\lsh}{\mathcal{L}_s(\hi)} %bounded sa operators
\newcommand{\sh}{\mathcal{S(H)}} %states
\newcommand{\ip}[2]{\left\langle\,#1\,|\,#2\,\right\rangle} %inner product
\newcommand{\ket}[1]{|#1\rangle} %ket
\newcommand{\bra}[1]{\langle#1|} %bra
\newcommand{\kb}[2]{|#1\rangle\langle#2|} %ketbra
\newcommand{\no}[1]{\left\|#1\right\|} %norm
\renewcommand{\tr}[1]{\textrm{tr}\left[#1\right]} %trace
\newcommand{\id}{\mathbbm{1}} %identity operator
\newcommand{\Ao}{\mathsf{A}}%generic observable
\newcommand{\Bo}{\mathsf{B}}%generic observable
\newcommand{\Co}{\mathsf{C}}%generic observable
\newcommand{\Jo}{\mathsf{J}}%jordan joint observable
\newcommand{\Mo}{\mathsf{M}}%generic observable
\newcommand{\No}{\mathsf{N}}%generic observable
\newcommand{\Po}{\mathsf{P}}%sharp observable
\newcommand{\Qo}{\mathsf{Q}}%sharp observable
\newcommand{\To}{\mathsf{T}}%trivial observable

\newcommand{\chan}{\mathfrak{C}}
\newcommand{\ch}[1]{\mathfrak{C}_{#1}}

\newcommand{\Cc}{\mathcal{C}} %channel
\newcommand{\Ec}{\mathcal{E}} %channel
\newcommand{\Ac}{\mathcal{A}} %channel

\newcommand{\Ii}{\mathcal{I}}

\newcommand{\Dev}{\mathfrak{D}} %device
\newcommand{\Triv}{\mathfrak{T}} %trivial device

\newcommand{\pleq}{\preceq}
\newcommand{\pgeq}{\succeq}
\newcommand{\obs}{\mathfrak{O}}

\begin{document}\setlength{\arraycolsep}{2pt}

\title[An Invitation to Quantum Incompatibility]{An Invitation to Quantum Incompatibility}

\author{
Teiko Heinosaari$^1$,
Takayuki Miyadera$^2$,
M\'ario Ziman$^3$
}
\address{$^1$Turku Centre for Quantum Physics, Department of Physics and Astronomy, University of Turku, FI-20014 Turku, Finland}
\address{$^2$Department of Nuclear Engineering, Kyoto University,
Kyoto daigaku-katsura, Nishikyo-ku, Kyoto, 615-8540, Japan}
\address{$^3$Institute of Physics, Slovak Academy of Sciences, D\'ubravsk\'a cesta 9, 84511 Bratislava, Slovakia}

\begin{abstract} 
In the context of a physical theory, two devices, A and B, described by the theory are called incompatible if the theory does not allow the existence of a third device C that would have both A and B as its components.
Incompatibility is a fascinating aspect of physical theories, especially in the case of quantum theory. 
The concept of incompatibility gives a common ground for several famous impossibility statements within quantum theory, such as ``no-cloning'' and ``no information without disturbance''; these can be all seen as statements about incompatibility of certain devices.
The purpose of this paper is to give a concise overview of some of the central aspects of incompatibility. 
\end{abstract}

%%%%%%%%%%%%%%%%%%%%%%
\section{Introduction}\label{sec:intro}
%%%%%%%%%%%%%%%%%%%%%%

The roots of quantum incompatibility go back to Heisenberg's uncertainty principle \cite{Heisenberg27} and Bohr's notion of complementarity \cite{Bohr28}. 
The basic lesson from those early studies is that there exist quantum measurements that cannot be implemented simultaneously, and it is in this sense that they are incompatible. 
At first sight, incompatibility of quantum measurements may seem more like an obstacle than an advantage. 
However, it has been realized that only incompatible measurements enable the violation of a Bell inequality \cite{Fine82}, \cite{MaAcGi06}, \cite{WoPeFe09}, one of the most intriguing phenomenon within the realm of quantum physics.
Bell inequalities are frequently used to prove the suitability of an experimental setting for quantum information processing tasks, such as quantum cryptography \cite{BrCaPiScWe14}.
This motivates attempting to see incompatibility as a useful resource and investigating its features and possible uses. 

It is illustrative to compare the incompatibility of pairs of measurements to the entanglement of bipartite states. 
There are various common features between incompatibility and entanglement. 
Both of these are non-classical properties, meaning that classical physical systems can possess neither property. 
Another common feature is that noise destroys both entanglement and incompatibility.
One can define several relevant notions analogously for both of these concepts:
for instance, an incompatibility breaking channel can be defined in an analogous way to an entanglement breaking channel \cite{HeKiReSc15}.
As with entanglement, incompatibility can be formulated and studied in the continuous variable Gaussian setting \cite{HeKiSc15}.
It is also possible to study the robustness of incompatibility in the same way as the robustness of entanglement \cite{Haapasalo15}.
There are many more connections that have either been studied or wait to be studied.
 
Incompatibility is traditionally thought of as a property of a collection of measurements, but the concept can be easily generalized to other collections of input-output devices. 
In this way, the concept of incompatibility gives a common ground for several famous impossibility statements within quantum theory, such as 'no-cloning' \cite{ScIbGiAc05} and 'no information without disturbance' \cite{Busch09},\cite{QI01Werner}; these can be all seen as statements about the incompatibility of certain devices. 
The generalized notion of incompatibility also opens up the investigation to families of more complex quantum devices, such as process measurements \cite{SeReChZi15}. 
This has the potential to reveal some new quantum limitations or applications.

Incompatibility can be defined equally well within a general operational theory as quantum theory.
Then it becomes clear that, as presumed, in a classical theory all devices are compatible.
More interestingly, it is possible to compare operational theories with respect to the maximal amount of incompatibility that they can host. 
Quantum theory does, in fact, contain pairs of devices that are as incompatible as a pair can be in any operational theory \cite{BuHeScSt13}, \cite{HeScToZi14}.
Investigating the fundamental features of quantum theory has a long tradition, and the parents of incompatibility, complementarity and uncertainty principle, have been studied within an axiomatic framework already some time ago \cite{BuLa80}.  
From this kind of foundational point of view incompatibility has not been  studied extensively, and there are many open question.
In particular, it would be interesting to see if the degree of incompatibility within quantum theory can be derived from some foundational principles.

The purpose of the present paper is to give a concise overview of some of the central aspects of incompatibility. In Section \ref{sec:incompatibility} we start with the general formulation of incompatibility for input-output devices in an operational theory. We discuss the quantification of incompatibility and some other general features that are most clearly stated at this general level. In Sections \ref{sec:observables} and \ref{sec:other} we concentrate on quantum devices, mostly on observables and channels, but we also point out elementary results for the incompatibility of process observables. 
In Section \ref{sec:order} we review the order theoretic characterization of quantum incompatibility.
Some final remarks are given in Section \ref{sec:outlook}.

%%%%%%%%%%%%%%%%%%%%%%%%%%%%%%%%%%%%%%
\section{Incompatibility in operational theories}\label{sec:incompatibility}
%%%%%%%%%%%%%%%%%%%%%%%%%%%%%%%%%%%%%%

%%%%%%%%%%%%%%%%%%%%%%%%%%%%%%%%%%%%%%
\subsection{Preliminary definition of incompatibility}\label{sec:prelim}
%%%%%%%%%%%%%%%%%%%%%%%%%%%%%%%%%%%%%%

\begin{figure}\begin{center}
\includegraphics[width=6cm]{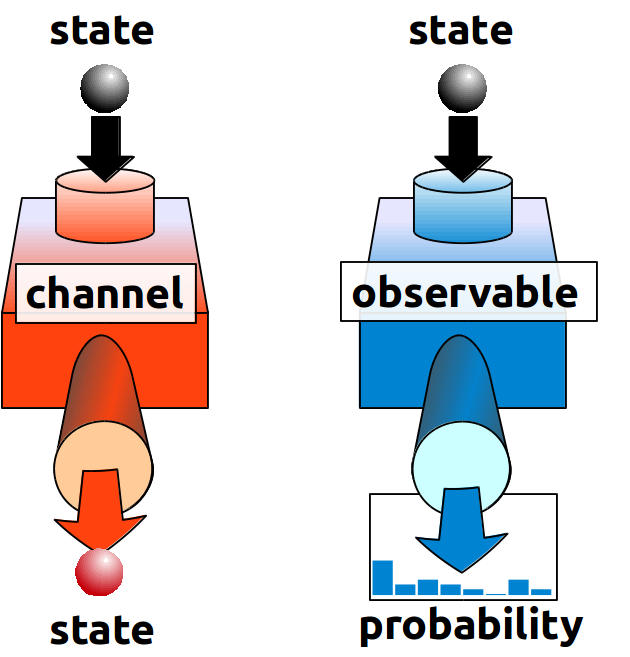}
\caption{A channel and observable illustrated as input-output devices.}
\label{fig:basic}
\end{center}
\end{figure}

Before we go into the mathematical definition of incompatibility, we shall try to grasp the concept intuitively.
We consider physical devices as boxes that have input and output ports.
For simplicity, we restrict to devices that have a single input port but possibly several output ports.  
An input for a device is taken to be a physical system, like a photon or neutron.
An output can be a transformed physical system, or a measurement outcome, or both.
For instance, an optical fibre has a photon in a polarization state as both an input and output.
This kind of device is called a channel.
A different kind of device is the one that gives a measurement outcome as an output; this is called an observable. 
For instance, in the quantum optical setting a physical implementation of an observable can be a combination of beam splitters, phase shifters and photo detectors.
We can still think of the whole setup as an input-output box. 
A channel and observable as input-output devices are illustrated in Fig. \ref{fig:basic}.

\begin{figure}\begin{center}
\includegraphics[width=10cm]{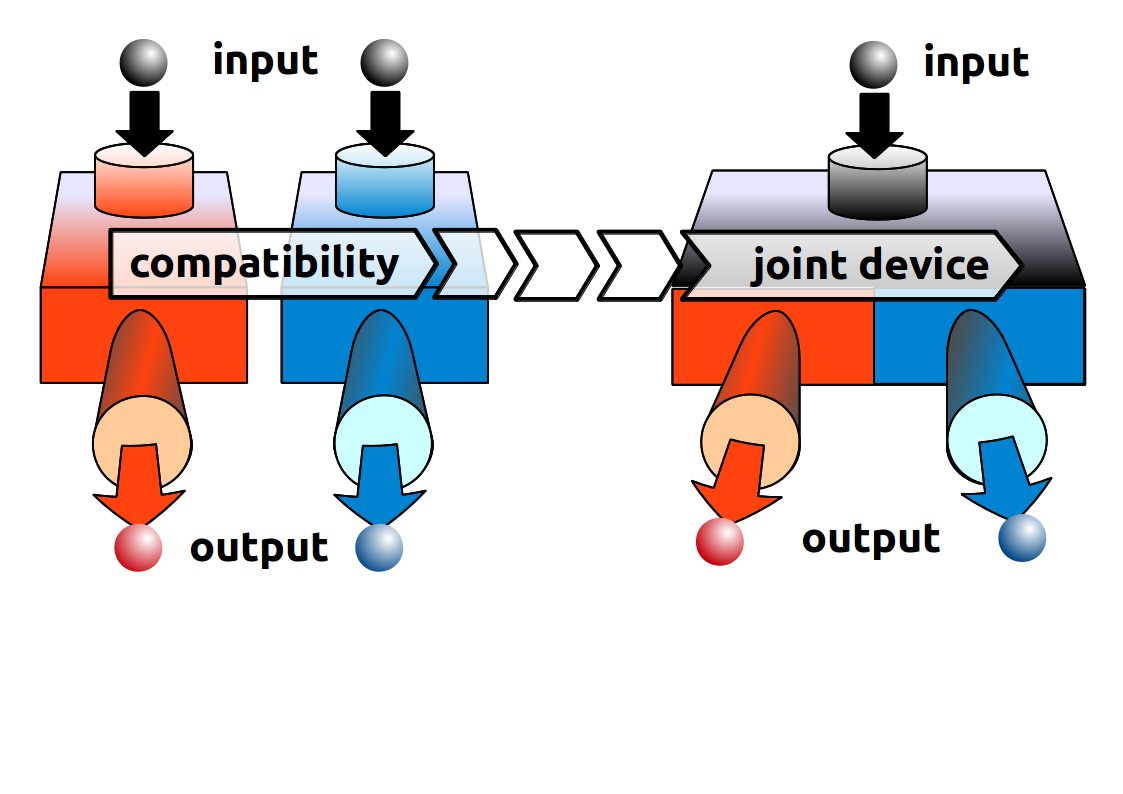}
\caption{Compatibility of two devices A (red) and B (blue) means that there exists a third device C that is their joint device.
Two devices are called incompatible if they don't allow a joint device.}
\label{fig:existence}
\end{center}
\end{figure}

Channels and observables with a single output port are the two most basic types of devices.
We may also have devices with several output ports.
As an example, a device may take a system as an input and then produce a measurement outcome together with a system in a transformed state. 
Whenever we have a multiport device, we can ignore all but one of the output ports and thus concentrate only on some part of the total device.
In this sense, a device with multiple output ports is a \emph{joint device} for two or more devices with one output ports.
We are now ready to state our preliminary definition of incompatibility by first stating its antonym, compatibility.
Two devices $A$ and $B$, both having a single output port, are called \emph{compatible} if there is a third device $C$ with two output ports such that $C$ is a joint device of $A$ and $B$; see Fig. \ref{fig:existence}.
If a joint device does not exist, then $A$ and $B$ are called \emph{incompatible}.

The essential point behind compatibility is that two devices that would by otherwise need a separate input each can be implemented by using just one input.  
Therefore, if a pair of devices is compatible, then one does not have to make a choice which one to implement, whereas with an incompatible pair of devices the choice is mandatory. In other words the incompatibility forces us to choose.

%%%%%%%%%%%%%%%%%%%%%%%%%%%%%%%%%%%%%%
\subsection{Incompatibility in operational theories}\label{sec:definition}
%%%%%%%%%%%%%%%%%%%%%%%%%%%%%%%%%%%%%%

In what follows we will put the previous preliminary definition of incompatibility into a proper mathematical context. For this purpose, we first briefly recall some basic elements of the framework called \emph{operational theory} or \emph{probabilistic theory} in the studies in quantum foundations; see e.g. \cite{BaBaLeWi07},\cite{ChDaPe10},\cite{StBu14} for further details.
Before we can give the definition of incompatibility, we need the concepts of a composite state space and a reduced state.

The basic ingredients of an operational theory are states and devices.
A state is a mathematical description of a preparation procedure of a system, while a device is a procedure applied after preparation. 
A device can operate only on certain kinds of systems, so systems can be understood as labels on input and output ports of devices. 
For each system described by the theory, there is a state space $\mathcal{S}$, which is assumed to be a convex subset of a real vector space $\mathcal{V}$, the convexity reflecting the possibility to mix preparations.

The state space of a \emph{classical system} is the set $\mathcal{P}(\Omega)$ of all probability distributions on a set $\Omega$. 
By a \emph{classical theory} we mean a theory that describes only classical systems.
An operational theory typically contains classical systems as special cases and then some non-classical systems that have state spaces of a different type.
In quantum theory the state space of a quantum system is identified with the set of all positive trace class operators of trace one on a complex Hilbert space $\hi$. 
The dimension of $\hi$ depends on the specific system.

Devices are, mathematically speaking, functions from one state space to another.
The basic requirement for all devices is that they are affine functions, i.e., a convex mixture of two inputs is mapped into the convex mixture of their outputs. 
The input and output state spaces of a device determine its type. 
An operational theory may have additional requirements for devices so that not all affine maps between two state spaces are valid devices. 
For instance, in quantum theory one has the requirement of complete positivity.

The simplest device on a state space $\mathcal{S}$ is an \emph{effect}, which is an affine function $e$ from $\mathcal{S}$ to the classical state space $[0,1]$. Physically speaking, an effect corresponds to a yes-no measurement that produces an outcome ``yes'' with the probability $e(s)$ and ``no'' with the probability $1-e(s)$. We denote by $\mathcal{E}(\mathcal{S})$ the set of all effects on $\mathcal{S}$. The \emph{unit effect} $u$ is the constant function $u(s)\equiv 1$.

An operational theory must specify the description of composite systems. 
Let $\mathcal{S}_1 \subset \mathcal{V}_1$ and $\mathcal{S}_2 \subset \mathcal{V}_2$ be two state spaces. 
With some reasonable assumptions the composite state space, denoted by $\mathcal{S}_1  \otimes \mathcal{S}_2$, 
can be identified with a convex subset of the tensor product vector space  $\mathcal{V}_1\otimes \mathcal{V}_2$. 
There is, however, not a unique choice \cite{NaPh69} and one has to understand the choice of the composite state space as a part of the definition of a specific operational theory.
We assume that the composite state space always contains the minimum tensor product $\mathcal{S}_1 \otimes_{min} \mathcal{S}_2$, which is the set of all convex combinations of the product elements $s_1\otimes s_2$ 
for $s_1\in\mathcal{S}_1$, $s_2\in\mathcal{S}_2$.
For classical systems this is the unique choice of a tensor product, but otherwise not.  
In quantum theory the composite state space is strictly larger than the minimum tensor. 
The convex combinations of product elements are referred to as separable states, and the other states are called entangled.

The composite system has an effect denoted by $e_1 \otimes e_2$ 
for each $e_1 \in \mathcal{E}(\mathcal{S}_1)$ and 
$e_2\in \mathcal{E}(\mathcal{S}_2)$, which represents independently applied measurements of 
$e_1$ and $e_2$. 
For a state $s\in\mathcal{S}_1  \otimes \mathcal{S}_2$, we define the \emph{reduced states} or \emph{marginal states} $marg_1(s)\in\mathcal{S}_1$ and $marg_2(s)\in\mathcal{S}_2$ via the conditions
\begin{equation}
e_1( marg_1(s)) = (e_1 \otimes u_2)(s) \, , \quad e_2(marg_2(s)) = (u_1 \otimes e_2)(s) \, ,
\end{equation}
required to hold for all effects $e_1\in \mathcal{E}(\mathcal{S}_1)$ and $e_2\in\mathcal{E}(\mathcal{S}_2)$, and where $u_1$ and $u_2$ are the unit effects in $\mathcal{E}(\mathcal{S}_1)$ and $\mathcal{E}(\mathcal{S}_2)$, and respectively.
The state $s$ is a \emph{joint state} of $marg_1(s)$ and $marg_2(s)$.
In quantum theory the marginal states of a state of a composite system are obtained by taking partial traces of the corresponding operator. 

The previous notions of marginal states and joint states are now lifted to devices. 
Let $\mathcal{S}$, $\mathcal{S}_1$ and $\mathcal{S}_2$ be state spaces and let us consider a device $\Dev:\mathcal{S}\to\mathcal{S}_1 \otimes \mathcal{S}_2$.
The \emph{marginals} of $\Dev$ are defined as
\begin{equation}\label{eq:marginals}
\Dev_1(s) := marg_1(\Dev(s)) \, , \quad \Dev_2(s) := marg_2(\Dev(s)) \, .
\end{equation}
The device $\Dev$ is a \emph{joint device} of $\Dev_1$ and $\Dev_2$.

We are then ready for the definition of our main concept.

\begin{definition*}
Two devices $\Dev_1: \mathcal{S} \to \mathcal{S}_1$ and 
$\Dev_2: \mathcal{S} \to \mathcal{S}_2$ with the same input space but possibly different output spaces are \emph{compatible} if there exists a device $\Dev: \mathcal{S} \to \mathcal{S}_1 \otimes \mathcal{S}_2$ such that $\Dev_1$ and $\Dev_2$ are the marginals of $\Dev$.
Otherwise $\Dev_1$ and $\Dev_2$ are \emph{incompatible}.
\end{definition*}

Let us remark that a compatible pair of devices need not have a unique joint device. The reason is simply that the marginal conditions \eqref{eq:marginals} specify the map $s \mapsto  \Dev(s)$ only partially. In particular, if two devices $\Dev_1$ and $\Dev_2$ are compatible and have two joint devices, then all convex mixtures of these  devices are also joint devices of $\Dev_1$ and $\Dev_2$. 
We conclude that a compatible pair of devices has either a unique joint device or an infinite number of them.  Let us also note that definitions of marginals and compatibility naturally extend to any finite set of devices. 

%%%%%%%%%%%%%%%%%%%%%%%%%%%%%%%%%%%%%%
\subsection{Quantification of incompatibility}\label{sec:degree}
%%%%%%%%%%%%%%%%%%%%%%%%%%%%%%%%%%%%%%

There are many ways to quantify the degree of incompatibility within a collection of incompatible devices. 
One can, for instance, start by defining a distance on the set of devices and then see how far the closest compatible devices are from the given incompatible collection.
However, what we discuss here a method that does not require a distance and is applicable to all devices, even of different type. 
This approach also allows us to compare incompatibility between different operational theories.

A device that gives a fixed output independently of the input is called a \emph{trivial device}.
For instance, a trivial observable corresponds to a coin tossing experiment, where the input state is ignored and the output is decided by tossing a coin. 
A trivial device is compatible with any other collection of devices.
This is obvious from a physical point of view, since the input state for a trivial device can be replaced with any fixed state.
In that sense, a trivial device does not need an input state and the input is therefore saved for another device that is desired to be implemented jointly.

\begin{figure}\begin{center}
\includegraphics[width=9cm]{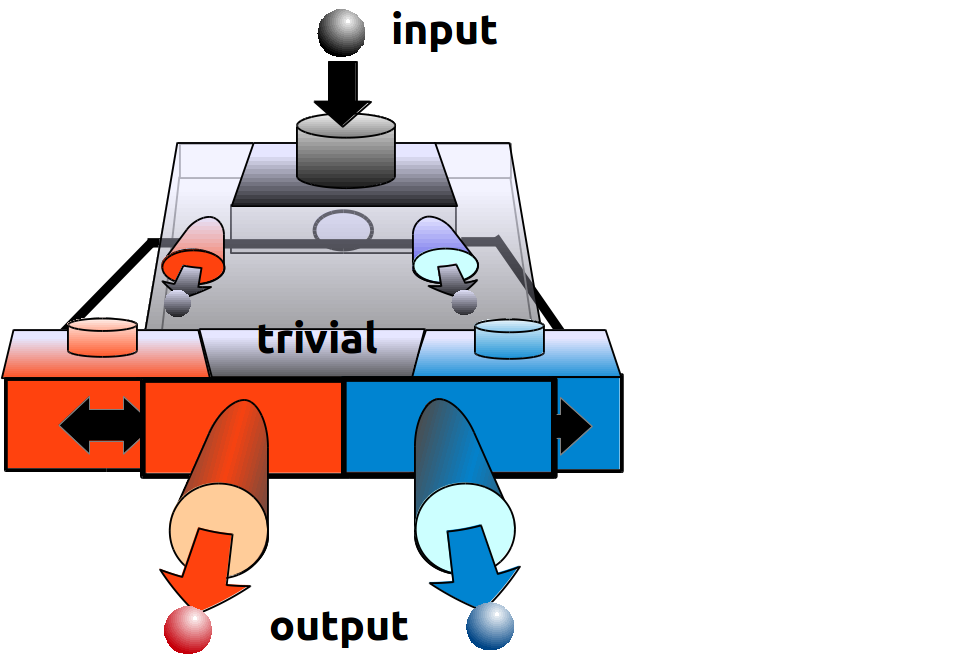}
\caption{
The idea of an approximate joint implementation of an arbitrary pair of devices, mixed with trivial ones, is illustrated. The coin toss decides whether the red or blue device is acting on the input state, while the remaining output is generated by a suitable trivial device. }
\label{fig:trivimix}
\end{center}
\end{figure}

Suppose we are considering an approximate implementation of $n$ incompatible devices $\Dev_1,\ldots,\Dev_n$.
For this purpose, we fix $n$ trivial devices $\Triv_1,\ldots,\Triv_n$, one of the same type for each device.
In each measurement run we roll an $n$-sided dice and, depending on the result, we implement one of the devices $\Dev_1,\ldots,\Dev_n$.
In addition to this, we implement $n-1$ trivial devices corresponding to those indices that were not chosen; see Fig. \ref{fig:trivimix}
We pretend that the outputs of the trivial devices are the outputs for the devices that were not implemented.
As a result, we have implemented $n$ devices $\Dev'_1,\ldots,\Dev'_n$ of the form
\begin{equation}\label{eq:mixtoss}
\Dev'_j = \tfrac{1}{n} \Dev_j + \tfrac{n-1}{n} \Triv_j \, ,
\end{equation}
and we can regard $\Dev'_j$ as a noisy version of $\Dev_j$.
It should be emphasized that this procedure works for all collections of $n$ devices as it only includes mixing and dice rolling.

This universal way of approximating incompatible devices with compatible ones motivates to look the best possible approximation of this form. 
Hence, for devices $\Dev_1,\ldots,\Dev_n$, we look for numbers $0\leq \lambda_j \leq 1$ such that there exist trivial devices $\Triv_1,\ldots,\Triv_n$ making the $n$ mixed devices 
$\lambda_j \Dev_j + (1-\lambda_j) \Triv_j$ compatible. 
The set of those points $(\lambda_1,\ldots,\lambda_n)\in [0,1]^n$ for which there exist such trivial devices is called the \emph{compatibility region} of the devices $\Dev_1,\ldots,\Dev_n$ \cite{BuHeScSt13},\cite{Gudder13}.
The compatibility region characterizes how much noise (in terms of trivial observables) we need to add to obtain compatible approximations.
If the devices $\Dev_1,\ldots,\Dev_n$ are compatible to start with, then their compatibility region is the hole hypercube $[0,1]^n$.

\begin{figure}\begin{center}
\includegraphics[width=5cm]{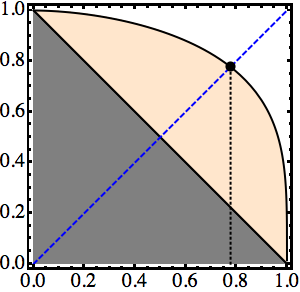}
\caption{The compatibility region (colored area) for a pair of observables illustrates how the addition of noise affects their compatibility. The triangle area (dark area) is in the compatible region of any pair of observables. 
The additional colored region (light are) is greater for more compatible pairs of observables. 
The degree of compatibility is obtained by taking the coordinate of the intersection point of the symmetry line (blue) and the boundary of the compatibility region.}
\label{fig:degree}
\end{center}
\end{figure}

It is sometimes desirable to quantify incompatibility with a single number.
For this purpose, we look at the subset of the compatibility region where the mixing parameters are equal, i.e.,  $\lambda_1=\cdots =\lambda_n \equiv\lambda$.
The \emph{degree of compatibility} of the devices $\Dev_1,\ldots,\Dev_n$, denoted by $deg(\Dev_1,\ldots,\Dev_n)$, is the supremum of numbers $0\leq \lambda \leq 1$ such that the $n$ devices 
$\lambda \Dev_j + (1-\lambda) \Triv_j$ are compatible for some choice of trivial devices $\Triv_1,\ldots,\Triv_n$ \cite{HeScToZi14}.
For instance, for two devices $\Dev_1$ and $\Dev_2$ the degree of compatibility is obtained as the intersection of the compatibility region and the symmetry line $\lambda_1=\lambda_2$; see Fig. \ref{fig:degree}.

\begin{figure}\begin{center}
\includegraphics[width=5cm]{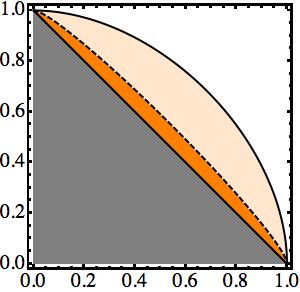}
\caption{The compatibility regions of finite dimensional position and momentum observables in dimensions $3$ (whole colored are) and $100$ (colored area bounded by the dashed line).
The compatibility region is the smaller the higher is the dimension.}
\label{fig:dim}
\end{center}
\end{figure}

As an example, let $\Qo_d$ and $\Po_d$ be the Fourier connected von Neumann observables on a finite $d$ dimensional Hilbert space, also called  finite dimensional position and momentum observables (see e.g. \cite{Vourdas97}).
It was shown in \cite{CaHeTo12} that their compatibility region is the set of those points $(\lambda_1,\lambda_2)\in [0,1]\times[0,1]$ that satisfy
\begin{equation}\label{eq:lambda}
(d-1) (\lambda_1 + \lambda_2) -  \sqrt{d  - (d-1) (\lambda_1 - \lambda_2)^2}\leq (d-2) \, .
\end{equation}
Hence, the degree of compatibility of $\Qo_d$ and $\Po_d$ is
\begin{equation}\label{eq:pos-mom}
deg(\Qo_d,\Po_d) = \half \left( 1 + \frac{1}{1+\sqrt{d}} \right) \, .
\end{equation}
Since the degree of compatibility of $\Qo_d$ and $\Po_d$ decreases as $d$ increases, it is justified to say that the finite dimensional position and momentum observables become more incompatible for increasing dimension $d$.
The compatibility regions in two cases are illustrated in Fig. \ref{fig:dim}.

As we concluded earlier, for any $n$ devices $\Dev_1,\ldots,\Dev_n$ we can form noisy compatible versions by tossing and mixing, and this leads to the devices given in \eqref{eq:mixtoss}.
This implies that the degree of compatibility of any collection of $n$ devices satisfies 
\begin{equation}
deg(\Dev_1,\ldots,\Dev_n) \geq \tfrac{1}{n} \, .
\end{equation}
For this reason, we say that a collection of $n$ devices $\Dev_1,\ldots,\Dev_n$ is \emph{maximally incompatible} if $deg(\Dev_1,\ldots,\Dev_n) = \tfrac{1}{n}$. 
The existence or non-existence of maximally incompatible devices should be seen as an intrinsic global property of an operational theory. 
A more refined question is whether an operational theory has maximally incompatible collections among some specific types of devices. 

Quantum theory does contain maximally incompatible pairs of observables. 
It was shown in \cite{HeScToZi14} that the standard position and momentum observables $\Qo$ and $\Po$ on the infinite dimensional Hilbert space $L^2(\R)$ are maximally incompatible, i.e., 
\begin{equation}\label{eq:pos-mom-true}
deg(\Qo,\Po) = \half \, .
\end{equation}
Another pair of complementary observables, namely the number and phase observables \cite{BuLaPeYl01}, was also shown to be maximally incompatible.
However, a pair of two-outcome quantum observables cannot be maximally incompatible, 
while in a different operational theory this is possible \cite{BuHeScSt13}.
We conclude that \emph{quantum theory contains maximally incompatible pairs of observables, but it does not include maximal incompatibility in the ultimate form}.
It seems to be is an open problem whether there exists a pair of finite outcome quantum observables which is maximally incompatible.

%%%%%%%%%%%%%%%%%%%%%%%%%%%%%%%%%%%%%%
\subsection{Broadcasting}\label{sec:broadcasting}
%%%%%%%%%%%%%%%%%%%%%%%%%%%%%%%%%%%%%%

\begin{figure}\begin{center}
\includegraphics[width=6cm]{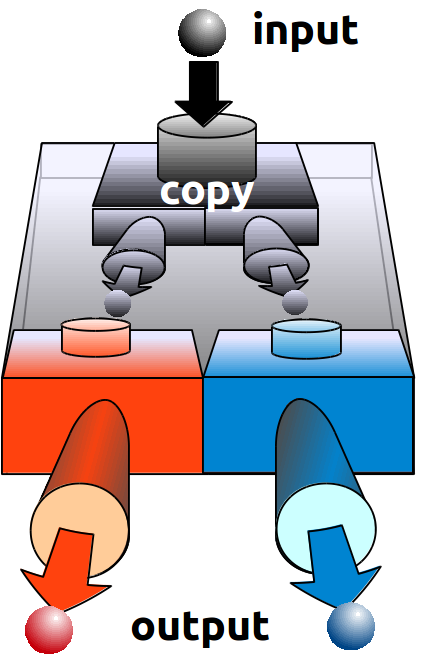}
\caption{A cloning device allows for joint implementation of arbitrary devices.}
\label{fig:cloning}
\end{center}
\end{figure}

If we are considering an operational theory where an unknown state can be copied, then any finite collection of devices is compatible. 
Namely, we can simply concatenate the desired devices $\Dev_1,\ldots,\Dev_n$ with the copying machine; see Fig. \ref{fig:cloning}.
The resulting device with multiple output ports is a joint device for $\Dev_1,\ldots,\Dev_n$.
This simple observation is more powerful that one would perhaps first expect; it implies that if a theory contains some incompatible devices, then a copying machine cannot exist in that theory \cite{QI01Werner}.

For a more detailed discussion of this concept, we recall that a device $\Ec:\mathcal{S}\to\mathcal{S}\otimes\mathcal{S}$ \emph{broadcasts} a state $\varrho$ if $marg_1 (\Ec(\varrho)) = \varrho$ and $marg_2(\Ec(\varrho)) = \varrho$.
(The difference to cloning is that the final state is not required to be the product state $\varrho\otimes\varrho$.)
A set of states is \emph{broadcastable} if there exists a channel $\Ec$ such that $\Ec$ broadcasts each state in that set.
It is known that a set of states is broadcastable if and only if it lies in a simplex generated by states that are distinguishable by a single-shot measurement \cite{BaBaLeWi07}.
It follows that broadcasting of all states is possible only in classical theories, whereas in quantum theory a set of states is broadcastable if and only if we restrict ourselves to a collection of mutually commuting density operators.

As universal broadcasting is impossible in quantum theory, its approximate versions have been investigated extensively \cite{ScIbGiAc05}.
From these studies we can infer some general limits on the degree of compatibility in quantum theory. 
In particular, a symmetric universal copying machine $C$ that makes $n$ approximate copies is of the form \cite{KeWe99}
\begin{equation}
C(\varrho) = s_{n,d} \, S(\varrho \otimes \id^{n-1}) S\, , 
\end{equation}
where $S$ is the projection from $\hi_d^{\otimes n}$ to the symmetric subspace of $\hi_d^{\otimes n}$ and the normalization coefficient $s_{n,d}$ does not depend on $\varrho$.
The state $\tilde{\varrho}$ of each approximate copy is obtained as the corresponding marginal of $C(\varrho)$ and, as it was shown in \cite{Werner98}, it reads
\begin{equation}
\tilde{\varrho} = c(d,n) \varrho + (1-c(d,n)) \frac{1}{d} \id \, ,
\end{equation}
where the number $c(d,n)$ is independent of $\varrho$ and given by
\begin{equation}
c(d,n) = \frac{n+d}{n(1+d)} \, .
\end{equation} 
An action of a device $\Dev$ on the transformed state $\tilde{\varrho}$ gives the same result as the action of the noisy device $c(d,n) \Dev + (1-c(d,n)) \Triv$ on the initial state $\varrho$, where $\Triv$ is the trivial device mapping all states into $\Dev(\frac{1}{d} \id)$.
Therefore, we conclude that the degree of compatibility has a lower bound
\begin{equation}\label{eq:bc-bound}
deg(\Dev_1,\ldots,\Dev_n) \geq \frac{n+d}{n(1+d)}
\end{equation}
for any choice of $n$ quantum devices on a $d$ dimensional quantum system.
It follows that maximally incompatible collections of devices, i.e., those having $deg(\Dev_1,\ldots,\Dev_n) = \frac{1}{n}$, do not exist in quantum theory if the dimension of the quantum system is finite \cite{HeScToZi14}.

It is an open question if the lower bound in \eqref{eq:bc-bound} is tight for quantum observables in the sense that there is equality for some observables $\Dev_1,\ldots,\Dev_n$. 
As observed in \cite{DaMaSa01}, for the usual complementary spin-$\half$ observables $\mathsf{X}$, $\mathsf{Y}$ and $\mathsf{Z}$ (corresponding to three mutual unbiased bases) the lower bound \eqref{eq:bc-bound} is not reached.
Namely, the lower bound in this case is $\frac{5}{9}$, but it follows from \cite{Busch86} that
$deg(\mathsf{X},\mathsf{Y},\mathsf{Z}) \geq \frac{1}{\sqrt{3}}$. 
The question of the most incompatible pair of quantum observables in a finite dimension $d$ seems to be open even in the simplest case of two-dimensional quantum systems.

%%%%%%%%%%%%%%%%%%%%%%%%%%%%%%%%%%%%%%
\subsection{Geometry of incompatibility}
%%%%%%%%%%%%%%%%%%%%%%%%%%%%%%%%%%%%%%

There is an extensive literature on the geometry of the state space (see e.g. \cite{GQS06}).
In particular, it is an interesting task to understand the boundary of entangled and separable states and the relative sizes of these sets.
This stimulates us to study similar questions in the context of compatible and incompatible pairs of devices, or more generally, collections of devices.

To formulate the geometric framework of incompatibility, we need to fix the types $T_1$ and $T_2$ of investigated devices.
The total set is then the Cartesian product $T_1\times T_2$.
We separate this set into the subset $K$ of compatible pairs and its complement set $K^C$ consisting of incompatible pairs.
The set $K$ is convex since a mixture of joint devices of two pairs gives marginals that are mixtures of the respective pairs. 
The separation of the total set into the sets of compatible and incompatible pairs is thus analogous to the separation of bipartite state space into separable and entangled states.

\begin{figure}\begin{center}
\includegraphics[width=6cm]{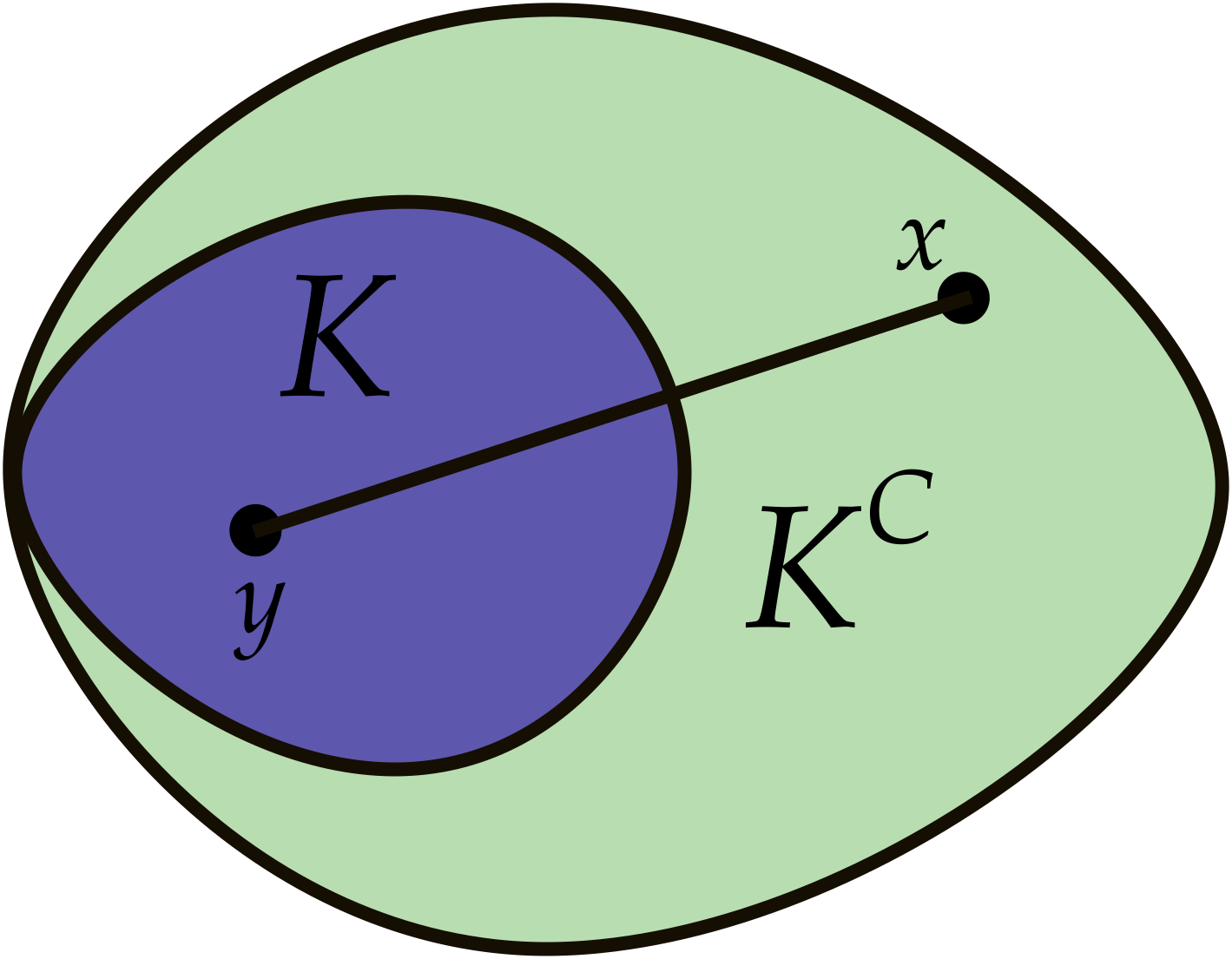}
\caption{The set of all pairs of devices is separated into the convex set $K$ of compatible pairs (darker area) and the complement set $K^C$ of incompatible pairs.
The boundary of $K$ and $K^C$ can be studied by investigating where line segments intersect it.}
\label{fig:geometry}
\end{center}
\end{figure}

The line segment between two points $x=(\Dev_1,\Dev_2)$ and $y=(\Dev'_1,\Dev'_2)$ in $T_1\times T_2$ consists of all pairs 
\begin{align}\label{eq:line}
\lambda x + (1-\lambda) y = (\lambda \Dev_1 + (1-\lambda) \Dev'_1,\lambda \Dev_2 + (1-\lambda) \Dev'_2) 
\end{align}
for $0\leq\lambda\leq 1$; see Fig. \ref{fig:geometry}.
To learn something about the structure of the set $K^C$, we can start with a fixed incompatible pair $x=(\Dev_1,\Dev_2)$ and then look for other pairs $y=(\Dev'_1,\Dev'_2)$ such that the line segment \eqref{eq:line} intersects the boundary of $K$ and $K^C$.
An illuminating task is to search for $y$ such that the weight $\lambda$ of $x$ for the intersection point is as short as possible. 
In finding the smallest possible weight $\lambda$ we can limit the choice of $y=(\Dev'_1,\Dev'_2)$, and there are at least three natural restrictions:
\begin{itemize}
\item[(a)] $\Dev'_1$ and $\Dev'_2$ are restricted to trivial devices
\item[(b)] $\Dev'_1$ and $\Dev'_2$ are restricted to compatible pairs of devices
\item[(c)] $\Dev'_1$ and $\Dev'_2$ can be any devices
\end{itemize}
Choice (a) is related to the degree of incompatibility that was discussed in Sec. \ref{sec:degree}, while (b) is, from the geometric point of view, analogous to the robustness of entanglement \cite{ViTa99}. 
This option was adopted in \cite{MaAcGi06} to quantify the degree of incompatibility.  
The third option (c) was studied recently in \cite{Haapasalo15}.
It was shown, for instance, that for a pair consisting of two unitary channels on a finite $d$ dimensional Hilbert space, the smallest weight $\lambda$ is
$\half \left( 1 + \frac{1}{d} \right)$,
while for a pair consisting of a von Neumann observable and a unitary channel the number is $\half \left( 1 + \frac{1}{\sqrt{d}} \right)$.
These numbers indicate again that higher dimensions permit greater amounts of incompatibility.  

%%%%%%%%%%%%%%%%%%%%%%%%%%%
\subsection{Operational compatibility vs descriptive compatibility}
%%%%%%%%%%%%%%%%%%%%%%%%%%%

Let us assume that two physicists, Alice and Bob, are using the same device but not necessarily simultaneously.
They may concentrate on different aspects or different functions of the device.
Since the origin of their description is the same device, their descriptions are necessarily consistent.
Assume, in contrast, that Alice and Bob deliver their descriptions to us without telling or possibly without even knowing that the origin for their descriptions is the same device.
It may happen that their descriptions are not consistent, meaning that there is not a single device that could give birth to both of their descriptions.
It is clear that this kind of consistency of descriptions is a precondition for compatibility.
However, it does not yet guarantee compatibility, since that would mean that the two devices can be simultaneously operated on a single input.

To further clarify this viewpoint, let us consider two devices $\Dev_1$ and $\Dev_2$, both with two output ports. 
If they are compatible, a joint device $\Dev$ for them would have four output ports according to our earlier definition.
One can also think of a device $\Dev'$ with only three output ports such that $\Dev_1$ is obtained when the first output port is ignored while $\Dev_2$ is obtained when the last output port is ignored. 
The essential difference between $\Dev$ and $\Dev'$ is that the latter uses the middle output port in both devices $\Dev_1$ and $\Dev_2$, while in the first one there is no such overlap.
When there is a need to distinguish these two situations, we say that our earlier definition of compatibility is \emph{operational compatibility} whereas this new notion is \emph{descriptive compatibility}.

\begin{figure}\begin{center}
\includegraphics[width=10cm]{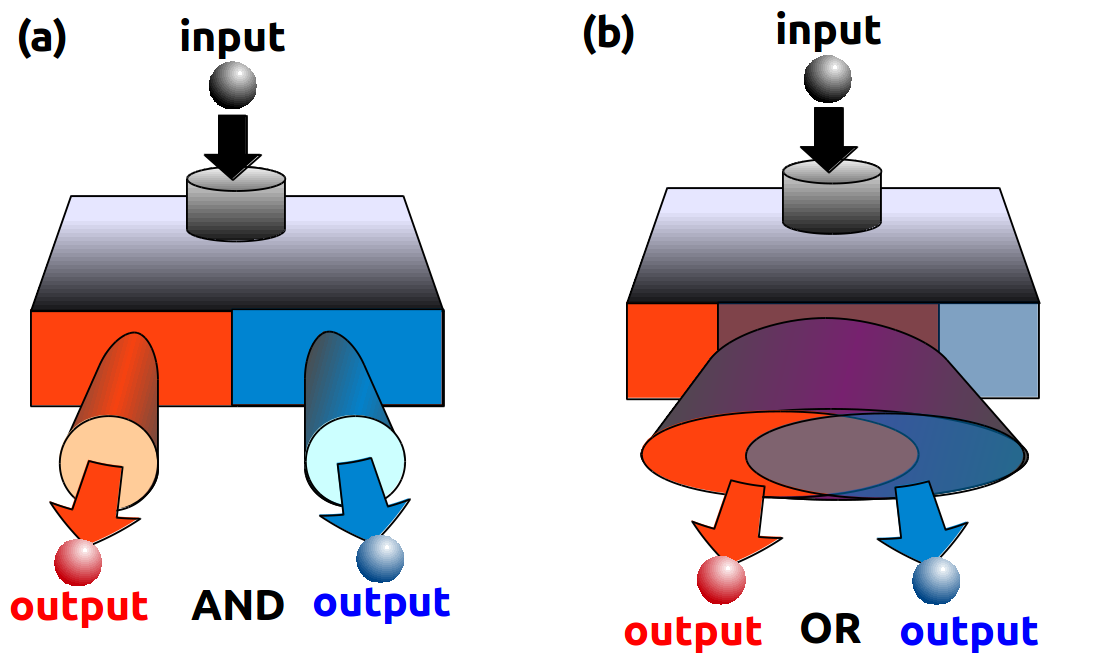}
\caption{Illustration of the difference between the concepts of
(a) operational and (b) descriptional compatibility.}
\label{fig:descriptional}
\end{center}
\end{figure}

The mathematical formulation of descriptive compatibility is similar to the definition of operational compatibility given in Subsec. \ref{sec:definition}, but now the marginals can have an overlap. 
This is exactly the reason why descriptive compatibility only means that the devices can be separately implemented on a single device, not necessarily simultaneously; see Fig. \ref{fig:descriptional}.
We observe that for observables the notions of operational and descriptive compatibility are equivalent since in that case we can get rid of any overlap simply by duplicating the obtained measurement outcomes.

The prototypical instance of the descriptive compatibility question is the \emph{state marginal problem} \cite{Klyachko04}. 
A simple form of the state marginal problem is the following: We have a composite system consisting of three systems $A,B,C$ and we are given two bipartite states $\varrho_1$ and $\varrho_2$ of subsystems $A\&B$ and $B \& C$, respectively. The question is: does there exists a state $\omega$ of the composite system $A\& B \& C$ such that $\varrho_1$ is the reduced state of the subsystem $A\&B$ and $\varrho_2$ is the reduced state of the subsystem $B \& C$? There are many variations of this question. 
For instance, often there is an additional requirement that the state $\omega$ of the composite system is pure. 
Let us stress again that the state marginal problem (with all its derivatives) is about \emph{consistency} of partial descriptions, not about their joint implementation.
The descriptive compatibility is relevant also in some other scenarios, including channel steering \cite{Piani15}.

%%%%%%%%%%%%%%%%%%%%%%%%%%%%%%%%%%%%%%
\section{Incompatibility of quantum observables}\label{sec:observables}
%%%%%%%%%%%%%%%%%%%%%%%%%%%%%%%%%%%%%%

%%%%%%%%%%%%%%%%%%%%%%%%%%%%%%%%%%%%%%
\subsection{Equivalent definitions of incompatibility for observables}\label{sec:functional}
%%%%%%%%%%%%%%%%%%%%%%%%%%%%%%%%%%%%%%

A quantum observable is a device that has a quantum input and classical output.
It is customary to use an equivalent mathematical description of a quantum observable as a \emph{positive operator valued measure} (POVM) \cite{OQP97}, \cite{MLQT12}.
A quantum observable with finite number of outcomes is identified with a function $x\mapsto \Mo(x)$ from the set of measurement outcomes $\Omega$ to the set $\lsh$ of selfadjoint operators on $\hi$.
For an input state $\varrho\in\sh$, the output probability distribution is $x \mapsto \tr{\varrho\Mo(x)}$.
Hence, the function $\Mo$ must satisfy $0\leq\Mo(x)\leq\id$ 
(operators satisfying this property are called \emph{effects}) 
for all $x\in\Omega$ and $\sum_{x\in\Omega} \Mo(x) = \id$. 
It is convenient to use the notation $\Mo(X) := \sum_{x\in X} \Mo(x)$ for any set $X\subset\Omega$.
The normalization condition then simply reads $\Mo(\Omega)=\id$. 
The observable $\Mo$ is called \emph{sharp} if for all $X\subset\Omega$ 
the effect $\Mo(X)$ is a projection, i.e. $\Mo(X)=\Mo(X)^*=\Mo(X)^2$. 
In other words, sharp observables correspond to projection-valued measures. 
We define the \emph{range of an observable} $\Mo$ as the set of effects $\Mo(X)$
associated with measurable sets $X\subset\Omega$.

Applying the general definition from Subsec.~\ref{sec:incompatibility} to the case of observables and translating it to the POVM language, we conclude that a joint observable is device that produces a list of outcomes $(x_1,x_2,\ldots,x_n)$ at each measurement round and its outcome set is a product set $\Omega_1\times\Omega_2\times\cdots\times\Omega_n$.
Ignoring all but the $k$th outcome determines an observable $\Mo_k$, given as
\begin{align}\label{eq:jointobs}
\Mo_k(x_k) = \sum_{l \neq k}\sum_{x_l} \Mo(x_1,x_2,\ldots,x_k,\ldots,x_n) \, .
\end{align}
As in the general case, we say that $\Mo_k$ is a \emph{marginal} of $\Mo$, and that $\Mo$ is a \emph{joint observable} of $\Mo_1,\ldots,\Mo_n$.
Hence, a finite collection of observables is compatible if and only if they are marginals of the same joint observable.
Traditionally, compatible observables are called \emph{jointly measurable}.
Using the set notation we can write \eqref{eq:jointobs} in the form
\begin{align}\label{eq:jointobs2}
\Mo_k(X) = \Mo(\Omega_1 \times \cdots \times \Omega_{k-1} \times X \times \Omega_{k+1} \times\cdots\times\Omega_n ) \, .
\end{align}
This way of writing the marginal condition is applicable also to observables with infinite number of outcomes; see e.g. \cite{LaYl04}.

There is another equivalent formulation of compatibility that may seem more intuitive. 
Suppose that we have a quantum observable $\Mo$ with an outcome space $\Omega$.
At each measurement round we get some outcome $x$. 
Since this is just a number, we can make as many copies of it as we want. 
We can further apply functions $f_1,\ldots,f_n$ to these copies, respectively.
We have thus produced $n$ outcomes $f_1(x),\ldots,f_n(x)$ even if we made only one measurement.
As a result, we have implemented $n$ observables $\Mo_1,\ldots,\Mo_n$, and we can write each of them as
\begin{align}\label{eq:function}
 \Mo_k(y) = \sum_{x} \delta_{y,f_k(x)}\Mo(x) \, ,
\end{align}
where $\delta_{a,b}$ is the Kronecker delta.
It is evident that the observable $\Mo$ implements simultaneously all the observables $\Mo_1,\ldots,\Mo_n$ even if $\Mo$ is not their joint observable in the strict sense of \eqref{eq:jointobs}.
In fact, \eqref{eq:jointobs} can be seen as a special case of \eqref{eq:function}, where the outcome space $\Omega$ is the product set $\Omega_1\times\Omega_2\times\cdots\times\Omega_n$ and each function $f_k$ is the projection map from $\Omega$ to $\Omega_k$. 
Let us note that this formulation is applicable to observables in any operational theory, and it has been called \emph{functional coexistence} \cite{LaPuYl98}.

The previous procedure can still be slightly generalized.
Again, we measure an observable $\Mo$ and start by making $n$ copies of the obtained measurement outcome $x$.
For each copy, we can have several possible relabeling functions.
If we obtain $x$, we toss a dice to decide which function we use to relabel the outcome.
We denote by $p_k(y\mid x)$ the conditional probability to relabel the $k$th copy of $x$ to $y$.
Hence, the actually implemented observables are given as
\begin{align}\label{eq:postobs}
\Mo_k(y) = \sum_{x} p_k(y \mid x) \Mo(x) \, .
\end{align}
Obviously, \eqref{eq:function} is a special instance of \eqref{eq:postobs}. However, \emph{if observables $\Mo_1, \ldots, \Mo_n$ can be written as in \eqref{eq:postobs} for some observable $\Mo$, then they are jointly measurable}. This can be seen as follows. 
From the observable $\Mo$ and the functions $p_k$ we define a new observable $\Mo'$ on the product set as
\begin{align}\label{eq:post->joint}
\Mo'(y_1,\ldots,y_n) =  \sum_x p_1(y_1 \mid x)\cdots p_n(y_n \mid x) \Mo(x) \, .
\end{align}
Then $\Mo'$ gives observables $\Mo_1, \ldots, \Mo_n$ as its marginals and is hence their joint observable.

Let us remark that an observable may have infinite number of outcomes.
In the case of countable infinite outcome sets, the previous argumentation still applies and the same conclusion on the equivalence of the three formulations is still valid.
If the outcome set $\Omega$ is uncountable, then one needs to specify a $\sigma$-algebra $\mathcal{F}$ consisting of subsets of $\Omega$ and observables must be literally consider as measures.
Proving the equivalence of the three formulations of joint measurability may require assumptions on measurable spaces $(\Omega,\mathcal{F})$ and the argumentation contains measure theoretic subtleties.
In the case where outcome sets are Hausdorff locally compact second countable topological spaces and $\sigma$-algebras are their Borel $\sigma$-algebras, the equivalence has been proven in \cite{AlCaHeTo09}.
Other equivalent formulations of joint measurability have been discussed in \cite{LaYl04}.

%%%%%%%%%%%%%%%%%%%%%%%%%%%%%%%%%%%%%%
\subsection{Commutativity and and its derivatives}\label{sec:comma}
%%%%%%%%%%%%%%%%%%%%%%%%%%%%%%%%%%%%%%

Traditionally, compatibility of observables has been often identified with their commutativity. The compatibility is, in fact, equivalent to commutativity for observables consisting of projections, i.e., for sharp observables. 
In the following we reproduce a proof of this fact and present a wider perspective on the role of commutativity.
A full list of equivalent conditions for the compatibility of sharp observables is presented in \cite{Lahti03}.  

We recall that the \emph{Jordan product} of two operators $E_1,E_2\in\lh$ is defined as $\half (E_1E_2+E_2E_1)$.
As a generalization, for each integer $n=2,3,\ldots$ we define a function $J_{n}:\lh^n\to\lh$ by
\begin{equation}\label{eq:jordan}
J_n(E_1,\ldots,E_n) := \frac{1}{n!} \sum_{\pi\in\Pi_n} E_{\pi(1)}\cdots E_{\pi(n)} \, ,
\end{equation}
where $\Pi_n$ is the set of all permutations of the set $\{1,2,\ldots,n\}$.
If $E_1,\ldots,E_n$ are selfadjoint operators, then also $J_n(E_1,\ldots,E_n)$ is a selfadjoint operator. 

The Jordan product can be used to define a joint observable.
Namely, let $\Mo_1,\ldots,\Mo_n$ be observables with an outcome space $\Omega$.
We define
\begin{equation}
\Jo(x_1,\ldots,x_n) := J_n(\Mo_1(x_1),\ldots,\Mo_n(x_n))
\end{equation}
for all $x_1,\ldots,x_n\in\Omega$.
Using \eqref{eq:jordan} we obtain
\begin{align}
\sum_{x_2,\cdots,x_n} \Jo(x_1,\ldots,x_n) = \Mo_1(x_1)
\end{align}
and similarly for other marginals.
This means that $\Jo$ is a joint observable whenever the operators $ \Jo(x_1,\ldots,x_n)$ are positive.
We thus obtain a sufficient condition for compatibility \cite{Heinosaari13}: \emph{observables $\Mo_1,\ldots,\Mo_n$ are compatible if the operator $J_n(\Mo_1(x_1),\ldots,\Mo_n(x_n))$ is positive for all $x_1,\ldots,x_n\in\Omega$}.

The fact that commuting observables are compatible is a consequence of the previous condition.
To see this, we observe that the product of $n$ commuting positive operators $E_1,\ldots,E_n$ is positive since 
\begin{equation}
E_1E_2\cdots E_n =  \left( \sqrt{E_1}\cdots\sqrt{E_{n}} \right)^* \left( \sqrt{E_1}\cdots\sqrt{E_{n}} \right) \, .
\end{equation}
It follows that $J_n(\Mo_1(x_1),\ldots,\Mo_n(x_n))$ is positive whenever the operators $\Mo_1(x_1),\ldots,\Mo_n(x_n)$ commute.
It should be noted that the previous sufficient condition for compatibility covers much wider class of compatible observables than just the commuting sets \cite{Heinosaari13}.

Let us then look the other side of the coin, namely, cases where non-commutativity is 
a sufficient criterion for incompatibility. 
Let $\Mo_1$ and $\Mo_2$ 
be two observables on a Hilbert space $\hi$. 
It was shown in \cite{MiIm08} that \emph{if operators $\Mo_1(x)$ and $\Mo_2(y)$ 
satisfy the inequality
\begin{align}\label{eq:tm-ineq}
\no{\Mo_1(x)\Mo_2(y) - \Mo_2(y)\Mo_1(x)} ^2 >
4 \no{ \Mo_1(x) - \Mo_1(x)^2 }
\cdot
\no{ \Mo_2(y) - \Mo_2(y)^2 } \, , 
\end{align}
then $\Mo_1$ and $\Mo_2$ are incompatible}. 
The number $\no{ \Mo_1(x) -\Mo_1(x)^2}$ quantifies the 
unsharpness of an effect $\Mo_1(x)$, and it vanishes if and only if 
$\Mo_1(x)$ is a projection. 
Therefore, as a special case of this result we see that if $\Mo_1(x)$ is a projection for some outcome $x$, then the compatibility of $\Mo_1$ and $\Mo_2$ requires that 
\begin{equation}
\Mo_1(x)\Mo_2(y)=\Mo_2(y)\Mo_1(x)
\end{equation}
 for all $y\in\Omega_2$.
 (For an alternative proof of this latter fact, see \cite{HeReSt08}.)

To explain the proof of the statement that \eqref{eq:tm-ineq} implies 
incompatibility, let us suppose that $\Mo_1$ and $\Mo_2$ are compatible 
observables. We denote their joint observable by $\Mo$. Due to the Naimark 
dilation theorem $\Mo$ can be presented as a restriction of a sharp 
observable on a larger Hilbert space; there exist a Hilbert space $\hik$, an isometry 
$V: \hi \to \hik$, and a sharp observable $\hat{\Mo}$ 
on $\hik$ satisfying 
\begin{equation}
V^*\hat{\Mo}(x,y)V = \Mo(x,y) \, .
\end{equation}
This sharp observable also defines Naimark dilations of $\Mo_1$ and $\Mo_2$ 
by $\hat{\Mo}_1(x):= \sum_y \hat{\Mo}(x,y)$ and $\hat{\Mo}_2(y) := \sum_x \hat{\Mo}(x,y)$.
Now, for two bounded operators $C$ and $D$ we have, by the $C^*$-property of the operator norm, 
\begin{equation}
\no{C^* D}^2 = \no{(C^*D)^*(C^*D)} = \no{D^*CC^*D} \, .
\end{equation}
Noting that $D^*C C^*D \leq \Vert C C^*\Vert D^* D$
we further get
\begin{equation} 
\Vert D^*C C^*D\Vert \leq 
\Vert CC^*\Vert \Vert D^*D\Vert 
= \Vert C^* C\Vert \Vert D^*D\Vert \, . 
\end{equation}
Using this operator norm inequality  
for $C= \sqrt{\id - VV^*} \hat{\Mo}_2(y) V$ and 
$D= \sqrt{\id - VV^*} \hat{\Mo}_1(x) V$ we then obtain 
\begin{align}
\Vert V^* \hat{\Mo}_1(x) \hat{\Mo}_2(y) V - \Mo_1(x)\Mo_2(y) \Vert 
\leq \Vert \Mo_1(x)  - \Mo_1(x)^2 \Vert^{1/2} 
\cdot \Vert \Mo_2(y) - \Mo_2(y)^2 \Vert^{1/2}.  
\end{align}
The observables $\hat{\Mo}_1$ and $\hat{\Mo}_2$ commute with each other as $\hat{\Mo}$ is commutative, and using this we get
\begin{eqnarray*}
&&
\Vert \Mo_1(x) \Mo_2(y) - \Mo_2(y) \Mo_1(x) \Vert 
\\
&=& \Vert
 (V^* \hat{\Mo}_2(y) \hat{\Mo}_1(x) V - \Mo_2(y)\Mo_1(x))  
-(V^* \hat{\Mo}_1(x) \hat{\Mo}_2(y) V - \Mo_1(x)\Mo_2(y)) \Vert 
\\
&\leq & 
\Vert
 V^* \hat{\Mo}_2(y) \hat{\Mo}_1(x) V - \Mo_2(y)\Mo_1(x) 
\Vert 
+ \Vert 
V^* \hat{\Mo}_1(x) \hat{\Mo}_2(y) V - \Mo_1(x)\Mo_2(y) \Vert 
\\
&\leq &
2 
\Vert \Mo_1(x)  - \Mo_1(x)^2 \Vert^{1/2} 
\cdot \Vert \Mo_2(y) - \Mo_2(y)^2 \Vert^{1/2} \, , 
\label{eq:tm-ineq-2}  
\end{eqnarray*}
and the claimed statement thus holds. 

%%%%%%%%%%%%%%%%%%%%%%%%%%%%%%%%%%%%%%
\subsection{Measurement uncertainty relations}\label{sec:ur}
%%%%%%%%%%%%%%%%%%%%%%%%%%%%%%%%%%%%%%

Starting from the famous article of W. Heisenberg \cite{Heisenberg27}, uncertainty relations have been studied extensively in many different variants.
Most of the uncertainty relations that can be found in the literature can be divided into preparation uncertainty relations and measurement uncertainty relations \cite{BuHeLa07},\cite{Muynck00}.
While preparation uncertainty relations are telling about the limitations how a quantum object can be prepared, measurement uncertainty relations set limitations on simultaneous measurements of two physical quantities.  
Many measurement uncertainty relations can be seen as necessary conditions for compatibility, or alternatively, as sufficient conditions for incompatibility. 
Reviewing the vast literature on uncertainty relations is beyond the scope of this paper. 
We will rather briefly exemplify their role as incompatibility tests.

The general setting for a measurement uncertainty relation is the following. 
We have two incompatible observables $\Mo$ and $\No$, and we have another pair of observables $\Mo'$ and $\No'$ which are consider as approximations of $\Mo$ and $\No$, respectively.
The qualities of these approximations are given by some 
nonnegative numbers $\delta(\Mo,\Mo')$ and $\delta(\No,\No')$. 
A measurement uncertainty relation is then a statement saying that if $\Mo'$ and $\No'$ are compatible, there should be a lower bound for some specified expression of $\delta(\Mo, \Mo')$ and $\delta(\No, \No')$, the lower bound obviously depending on $\Mo$ and $\No$.
In the simplest case the lower bound can be for the product $\delta(\Mo,\Mo')\cdot\delta(\No,\No')$ or the sum $\delta(\Mo,\Mo')+\delta(\No,\No')$, but it can be also for some more involved expression. 

As an example, the discrepancy between two observables $\Mo$ and $\Mo'$ can be quantified as
\begin{equation}
\delta(\Mo, \Mo') 
= \max_x \Vert \Mo(x) -\Mo'(x)\Vert 
= \max_x \sup_{\rho} \left| \mbox{tr}[\rho\Mo(x)] 
- \mbox{tr}[\rho \Mo'(x)]\right| \, , 
\end{equation}
while the inherent unsharpness of $\Mo$ can be quantified as
\begin{equation}
\nu(\Mo)=\max_x \Vert \Mo(x) -\Mo(x)^2\Vert \, .
\end{equation}
It was proved in \cite{MiIm08} that for these measures, the following inequality holds:
\begin{equation*}
2 \delta(\Mo, \Mo') 
\delta(\No, \No') + 
\delta(\Mo, \Mo') + \delta(\No, \No') 
+2 (2 \delta(\Mo, \Mo') + 
\nu(\Mo))^{1/2} 
(2 \delta(\No, \No') + \nu(\No))^{1/2} 
\geq c_{\Mo,\No} \, ,  
\end{equation*}
where 
\begin{equation}
c_{\Mo,\No} =  \max_{x,y} \Vert  \Mo(x)\No(y) - \No(y)\Mo(x) \Vert \, . 
\end{equation}
If we now fix $\Mo$ and $\No$, then the violation of this inequality is a sufficient condition for incompatibility of any two observables $\Mo'$ and $\No'$.

As an example, let us choose $\Mo=\Qo_d$ and $\No=\Po_d$, the finite dimensional position and momentum observables introduced in Subsec.~\ref{sec:degree}.
In this case $\nu(\Qo_d) = \nu(\Po_d)=0$ and $c_{\Qo_d,\Po_d} = \frac{\sqrt{d-1}}{d}$.
Thus we obtain 
\begin{equation}\label{eq:ur}
2 \delta(\Qo_d, \Mo') 
\delta(\Po, \No') + 
\delta(\Qo_d, \Mo') + \delta(\Po_d, \No') 
+4 \delta(\Qo_d, \Mo')^{1/2}  
\delta(\Po_d, \No')^{1/2}  
\geq \frac{\sqrt{d-1}}{d} \, , 
\end{equation} 
which holds for all compatible observables $\Mo'$ and $\No'$. 
Therefore, if two observables  $\Mo'$ and $\No'$ violate \eqref{eq:ur}, then they must be incompatible. 

%%%%%%%%%%%%%%%%%%%%%%%%%%%%%%%%%%%%%%
\subsection{Information and incompatibility}\label{sec:ic}
%%%%%%%%%%%%%%%%%%%%%%%%%%%%%%%%%%%%%%

An observable $\Mo$ is called \emph{informationally complete} if it gives different measurement outcome distributions to all quantum states \cite{Prugovecki77}, \cite{BuLa89}. 
In that way, an informationally complete observable allows the reconstruction of an unknown input state. 
Once we know the input state, we can calculate the probability distributions of any observable we want.
This may lead to a false thought that incompatibility can be circumvented by measuring an informationally complete observable.
It is important and instructive to understand that the existence of an informationally complete observable does not mean that all observables are jointly measurable.
To see the difference to joint measurements, we recall that (in a finite dimensional Hilbert space) an observable $\Mo$ is informationally complete if and only if any observable $\No$ can be written as a sum
\begin{equation}
\No(y) = \sum_x f_\No(x,y) \Mo(x) \, ,
\end{equation}
where $f_\No$ is a real valued processing function \cite{DaPeSa04}. 
This differs from \eqref{eq:postobs} since $f_\No$ can take negative values.
In practice, this means that if we measure an informationally complete observable $\Mo$ only once and obtain a single outcome, we cannot infer much on the outcomes of other observables.

An observation related to the difference between joint observables and informationally complete observables is that \emph{there are compatible observables that cannot have an informationally complete joint observable}.
Namely, suppose that $\Mo$ is a joint observable of some set of observables containing an observable $\Ao$ such that one of the operators, say $\Ao(1)$, is a projection. (For instance, $\Ao$ can be the observable that corresponds to an orthonormal basis $\{\varphi_x\}$, i.e., $\Ao(x)=\kb{\varphi_x}{\varphi_x}$.)
Since $\Ao(1)$ is a sum of some elements $\Mo(x_1,\ldots,x_n)$, it is clearly in the range of $\Mo$.
But an informationally complete observable cannot have a projection in its range \cite{BuCaLa95}, hence $\Mo$ is not informationally complete.
This example is, in fact, linked to the well-known foundational feature of quantum theory that it is impossible to identify an unknown quantum state if only a single system is available.
Namely, the observable $\Ao$ defined above has the property that if we know that the input state is one of the vector states $\varphi_x$ but we don't know which one, then $\Ao$ can determine the correct state already from a single outcome. 
Suppose that another observable $\Bo$ is informationally complete, hence able to identify a completely unknown state from the full measurement statistics. 
Then $\Ao$ and $\Bo$ are necessarily incompatible, as their hypothetical joint observable would be capable of performing both tasks, which is impossible by the earlier argument. 
It is also interesting to note that in some odd dimensions optimal approximate position and momentum observables allow informationally complete joint measurements, but in even dimensions not \cite{CaHeTo12}. 
In the infinite dimensional case there seems to be no connection between the informational completeness of a joint phase space observable and state distinction properties of its marginal observables \cite{Schultz12}.

Clearly, a joint observable of a set of observables gives at least as much information as each marginal observable. 
This leads to the idea that a set of observables must be incompatible if their hypothetical joint measurement would provide too much information. 
For make this idea useful, one has to formulate the concept of information in a proper way. 
As it was demonstrated in \cite{WaSaUe11}, \cite{Zhu15}, the Fisher information is a useful measure of information to make the intuitive idea to work. 
The method is then to use the quantum estimation theory to derive limitations on joint measurements.
 The limitations derived in \cite{Zhu15} for compatibility are particularly effective for multiple observables, in which case incompatibility conditions are less studied. For instance, let $\Mo_1, \ldots,\Mo_n$ be complementary von Neumann observables and let $\To$ be the trivial observable $\To(x)=\frac{1}{d} \id$.  It was proved in \cite{Zhu15} that the mixtures $\lambda_1 \Mo_1 + (1-\lambda_1) \To,\ldots,\lambda_n \Mo_n + (1-\lambda_n) \To$ are incompatible if
\begin{equation}
\sum_{j=1}^n \lambda_j^2  > 1 \, .
\end{equation}
This inequality is known also to be a necessary condition for incompatibility in the case of two or three complementary qubit observables \cite{Busch86}.

%%%%%%%%%%%%%%%%%%%%%%%%%%%%%%%%%%%%%%
\subsection{Coexistence}
%%%%%%%%%%%%%%%%%%%%%%%%%%%%%%%%%%%%%%

Deciding, either numerically or analytically,  
whether observables $\Mo_1$ and $\Mo_2$ are
jointly observable becomes more and more tedious as the number of 
outcomes increases. Therefore, any reduction of the
compatibility problem to a simplified compatibility problem 
is of general interest. 
On the more conceptual side, one may wonder if the joint measurability is essentially a property of the operators in the range of two observables or if it depends in the specific way in which an observable assigns an operator to an outcome. For instance, the commutativity of two observables is decided solely on the level of operators. 

Suppose $\Mo$ is an observable with an outcome set $\Omega$, and fix 
a subset $X\subset\Omega$. We may define a new (binary) observable 
$\Mo^X$ with the outcomes $1$ 
and $0$ as
\begin{equation}
\Mo^X(1) := \Mo(X) \, , \quad \Mo^X(0) := \id-\Mo(X) \, .
\end{equation} 
This observable is called a \emph{binarization of} $\Mo$.
One could expect that having the collection of all these binarizations 
should be, in some sense, same as having $\Mo$. From the compatibility 
point of view, we may introduce the following two natural concepts. 
Observables $\Mo_1,\ldots,\Mo_n$ are called
\begin{enumerate}[(a)]
\item \emph{coexistent} if the collection of all their binarizations $\Mo_\ell^{X_\ell}$ is jointly measurable.
\item \emph{weakly coexistent} if for any fixed choice of subsets $X_j \subseteq \Omega_j$, the collection of binarizations $\Mo_1^{X_1},\ldots,\Mo_n^{X_n}$ is jointly measurable.
\end{enumerate}
It follows immediately from these definitions that for a finite collection of observables, we have the following hierarchy of the properties:
\begin{align*}
\textrm{jointly measurable} \quad \Rightarrow \quad \textrm{coexistent} \quad \Rightarrow \quad \textrm{weakly coexistent} \, .
\end{align*}
It is also clear that the three concepts are equivalent for any collection of two-outcome observables.

The concepts of coexistence and joint measurability where clearly distinguished in \cite{LaPu97}, \cite{LaPu01} and it was noted that their equivalence or inequivalence is an open question. 
By providing suitable examples it was shown in \cite{HeReSt08} that 
\begin{align*}
\textrm{weakly coexistent}  \quad \nRightarrow \quad  \textrm{jointly measurable}
\end{align*}
and later in \cite{ReReWo13} a stronger result that
\begin{align*}
\textrm{coexistent} \quad \nRightarrow \quad \textrm{jointly measurable} \, .
\end{align*}
Finally, an example demonstrating that 
\begin{align*}
\textrm{weakly coexistent} \quad \nRightarrow \quad \textrm{coexistent}
\end{align*}
was given in \cite{HaPeUo15}.
Hence, the three concepts are indeed different. Let us note that all the relevant examples were using qubit observables and can be extended to higher dimensions, so the concepts are inequivalent in all dimensions.
However, even if the three concepts are inequivalent, there are some important classes of observables under which all three concepts coincide. For example, a pair of observables such that at least one of them is discrete and extreme in the convex set of observables is jointly measurable if and only if the pair is coexistent \cite{HaPeUo15}. 

%%%%%%%%%%%%%%%%%%%%%%%%%%%%%%%%%%%%%%
\subsection{Role of incompatibility in bipartite settings}
%%%%%%%%%%%%%%%%%%%%%%%%%%%%%%%%%%%%%%

The violation of local realism, demonstrated by the Einstein-Podolski-Rosen paradox \cite{EiPoRo35}, or the violation of Bell inequalities {\cite{Bell64}, is probably the most puzzling feature of quantum systems.
For this phenomenon to occur the existence of \emph{quantum entanglement} 
is essential, but not sufficient \cite{Werner89}. 
A requirement on the side of measurement devices is that the entangled 
particles must be probed by incompatible observables. 
In other words, without incompatibility we couldn't experience quantum nonlocality. 
A connection between joint measurements and Bell inequalities was investigated in \cite{SoAnBaKi05}. 
A tight relation between these two notions was proved in \cite{WoPeFe09}, where it was shown that an arbitrary pair of incompatible binary observables enables the violation of the Bell-CHSH inequality.
Further, violations of certain scaled versions of the Bell-CHSH inequality are related to some operationally motivated incompatibility monotones \cite{HeKiRe15}.
A compatible set of observables cannot violate any Bell inequality, but it still seems to be open question if there are incompatible sets that cannot violate any Bell inequality.
There is, however, some indication that measurement incompatibility would not imply Bell nonlocality \cite{QuBoHiBr15}.

Back in the 1930s, quite simultaneously with the discussion of the EPR paradox, 
E. Schr\"odinger realized \cite{Schrodinger35} that for bipartite quantum 
systems one of the parties can steer distantly the properties of the second
system by acting locally on his/her system and communicating
the classical information. 
He discovered the phenomenon that is now called \emph{quantum 
steering} \cite{WiJoDo07}. 
In quantum steering two parties, Alice and Bob, share a state $\omega_{AB}$. 
Suppose that Alice chooses to measure an observable $\Ao$ on her part of the composite system. 
After observing the outcome $x$ with the probability ${\rm tr}[(\Ao(x)\otimes I)\omega_{AB}]$, 
Bob's system is described by the conditional (subnormalized)
state 
\begin{equation}
\varrho_{B}^{x|\Ao}={\rm tr}_A [(\Ao(x)\otimes \id)\omega_{AB}]
\end{equation}
This conditional state does not depend on the specific way how Alice measures the observable $\Ao$.
Moreover, the ensemble on Bob's side is described by the same average state
\begin{equation}
\sum_x \varrho_B^{x|\Ao}=\varrho_B\equiv{\rm tr}_B[\omega_{AB}] \, , 
\end{equation}
for all choices of $\Ao$.
It is the decomposition of $\varrho_B$ that is the subject of Alice's 
steering. 
But can Alice really prove to Bob that she can affect his system?

Clearly, Alice has to send the choice and the result of her measurement 
to Bob, so that Bob can verify the conditional states. However, for any chosen
family of observables $\Ao_1,\dots,\Ao_n$ Bob could think 
of his system as being described by a collection of states $\varrho_\lambda$ 
distributed according to some (unknown) probability distribution $\pi_\lambda$ and in this way forming the marginal state $\varrho_B$. 
If there exist a valid conditional probability distribution $q(x|\Ao_j,\lambda)$, where $x$ labels potential 
outcomes of a measurement $\Ao_j$, such that
\begin{equation}
\varrho_B^{x|j}=\sum_\lambda \pi_\lambda q(x|\Ao_j,\lambda) \varrho_\lambda\,,
\end{equation}
for all observables $\Ao_j$ and all outcomes $x$, then Bob can provide a local explanation for the update of his state. 
If this is the case
then the action of Alice is not necessary in order to manipulate Bob's 
system into the post-selected state $\varrho_B^{x|j}$, and, consequently,
Alice cannot prove she is really \emph{steering} the state of Bob's system
distantly. 
Instead, she could cheat by preparing the ensemble 
$\{\pi_\lambda,\varrho_\lambda\}$ and sending the information in accordance
with probability $q(x|\Ao_j,\lambda)$ to prepare the desired states 
$\varrho_B^{x|j}$. 
It was shown in \cite{WiJoDo07} that entanglement 
of $\omega_{AB}$ is necessary to exhibit the quantum steering 
and also that this phenomenon is different from Bell's nonlocality.

We will refer to the set of subnormalized states
$\{\varrho^{x|j}\}$ as assemblage. 
The description of Bob's assemblage by an ensemble $\{\pi_\lambda,\varrho_\lambda\}$ and some 
conditional probabilities $q(x|\Ao_j,\lambda)$ is called 
a local hidden state model. The assemblage is then
called \emph{steerable} only if such local hidden state 
model does not exist.
It was shown in \cite{QuVeBr14,UoMoGu14} that
for the assemblage $\{\varrho^{x|j}\}$ associated with a family of observables
$\Ao_1,\dots,\Ao_n$ is steerable if and only if the observables are jointly measurable. 
In other words, the incompatibility of observables
is necessary and sufficient to demonstrate the phenomenon of quantum 
steering. 
This connection holds, in fact, in a general class of probabilistic theories \cite{Banik15}.

%%%%%%%%%%%%%%%%%%%%%%%%%%%%%%%%%%%%%%
\section{Incompatibility of other quantum devices}\label{sec:other}
%%%%%%%%%%%%%%%%%%%%%%%%%%%%%%%%%%%%%%

%%%%%%%%%%%%%%%%%%%%%%%%%%%%%%%%%%%%%%
\subsection{Incompatibility of quantum channels}
%%%%%%%%%%%%%%%%%%%%%%%%%%%%%%%%%%%%%%

A quantum channel is an input-output device that transforms quantum states into quantum states. The dimension of the output system need not be the same as the dimension of the input system, as a channel may, for instance, incorporate a new system. 

Let us consider a channel $\Cc$ that acts on states of a composite system $A+B$ consisting of two subsystems $A$ and $B$. The system $A$ can be in an arbitrary state $\varrho$, while the system $B$ is assumed to be in a fixed blank state $\varrho_0$. 
In this way we have a device with a single input port and two output ports.
After the channel has operated on the joint state $\varrho \otimes \varrho_0$, we isolate the subsystems and investigate their reduced states.
The overall procedure thus determines two channels 
\begin{align}\label{eq:channels}
\Cc_A:\varrho \mapsto {\rm tr}_B[\Cc(\varrho \otimes \varrho_0)] \quad \textrm{and} \quad  \Cc_B:\varrho \mapsto {\rm tr}_A[\Cc(\varrho \otimes \varrho_0)] \, .
\end{align}
Within the framework of incompatibility we are interested on the reverse question: if two channels $\Cc_A$ and $\Cc_B$ are given, is there a channel $\Cc$ acting on the composite system $A+B$ such that $\Cc_A$ and $\Cc_B$ are of the form \eqref{eq:channels} for some blank state $\varrho_0$? 
This is equivalent of asking if $\Cc_A$ and $\Cc_B$ are compatible in the sense of the general definition discussed in Subsec. \ref{sec:definition}.
In fact, suppose that $\Cc_A$ and $\Cc_B$ are compatible. 
This means that there exists a channel $\Ec$ such that 
$\mbox{tr}_B [ \Ec (\varrho)] = \Cc_A(\varrho)$ and 
$\mbox{tr}_A [\Ec (\varrho)] =\Cc_B(\varrho)$. 
Then according to the Stinespring dilation theorem
there exist an additional system $C$ and 
isometry $V: \hi_A \to \hi_A \otimes \hi_B \otimes \hi_C$ 
such that $\Ec(\varrho) 
= V \varrho V^*$ holds.  
This isometry can be extended 
to a unitary operator $U$ acting on $\hi_A \otimes \hi_B \otimes \hi_C$
and satisfying 
\begin{equation}
V \psi = U (\psi  \otimes \phi_0 \otimes \phi_1)
\end{equation} 
for all $\psi \in \hi_A$ and a fixed $\phi_0 \otimes \phi_1 \in \hi_B \otimes \hi_C$.
We define $\varrho_0 := |\phi_0\rangle \langle \phi_0 |$ 
and a joint device then reads 
\begin{equation}
\Cc(\varrho \otimes \varrho_0) = \mbox{tr}_{C}[ U
 (\varrho \otimes \varrho_0 \otimes |\phi_1\rangle \langle \phi_1| ) U^*] \, .
 \end{equation}

The most prominent example of incompatibility of quantum channels is related to the celebrated \emph{no-cloning theorem}  \cite{Dieks82},\cite{WoZu82}. 
We already noticed a connection between compatibility and cloning in Subsec. \ref{sec:broadcasting}, and now we point out a supplementing aspect. A \emph{universal cloning device} \cite{BuHi96} is a machine that accepts an unknown state $\varrho$ on its input and produces a state $\varrho\otimes\varrho$ on its output. We can relax this condition by not requiring the output state to be a product state but only demanding that its reduced states are exact copies of $\varrho$, and this is often called broadcasting in order to distinguish it from a cloning producing independent copies. 
If $\Cc$ is the broadcasting channel, then the broadcasting requirement is equivalent to the condition that both $\Cc_A$ and $\Cc_B$ in \eqref{eq:channels} are identity channels, i.e., $\Cc_A(\varrho)=\Cc_B(\varrho)=\varrho$ for all states $\varrho$. The question on the existence of universal broadcasting is then equivalent to the question of compatibility of these two identity channels. 
The fact that universal broadcasting is not possible in quantum theory is thus equivalent with the statement that two identity channels are incompatible. 

The incompatibility of two identity channels also serves as a demonstration that \emph{a channel need not be compatible with itself}. This is a clear difference compared to observables, and the underlying reason is, indeed, that classical information can be copied but quantum information cannot. 
To further illustrate incompatibility of channels, let us introduce a family of diagonalizing channels. 
These channels completely destroy quantum coherences and are not able to transfer quantum information at a nonvanishing rate, however, they keep orthogonality of exactly $d$ quantum states (forming an orthonormal basis), thus, they may act as noiseless from the point of view of classical information transfer. 
For each orthonormal basis $\mathcal{B}=\{\varphi_j\}_{j=1}^d$, we define a channel ${\rm diag}_{\mathcal{B}}$ as
\begin{equation}
{\rm diag}_{\mathcal{B}}(\varrho)=\sum_{j=1}^d \ip{\varphi_j}{\varrho\varphi_j}
\kb{\varphi_j}{\varphi_j}\, .
\end{equation}
Suppose that two diagonalizing channels ${\rm diag}_{\mathcal{B}}$ 
and ${\rm diag}_{\mathcal{B^\prime}}$ are  compatible. 
Then, by applying their joint channel $\Cc$, we can produce states  ${\rm diag}_{\mathcal{B}}(\varrho)$ and ${\rm diag}_{\mathcal{B^\prime}}(\varrho)$ from an input state $\varrho$.
By measuring in the basis $\mathcal{B}$ for the state ${\rm diag}_{\mathcal{B}}(\varrho)$ and in the basis $\mathcal{B^\prime}$ for the state ${\rm diag}_{\mathcal{B^\prime}}(\varrho)$, we are implementing a joint measurement of sharp observables corresponding to $\mathcal{B}$ and $\mathcal{B^\prime}$.
As we have seen in Subsec.~\ref{sec:comma} these two observables are compatible only if they commute. 
Therefore, we conclude that the diagonalizing channels ${\rm diag}_{\mathcal{B}}$ and ${\rm diag}_{\mathcal{B^\prime}}$ are  incompatible if $\mathcal{B} \neq \mathcal{B^\prime}$.

%%%%%%%%%%%%%%%%%%%%%%%%%%%%%%%%%%%%%%
\subsection{Conjugate channels}\label{sec:complementarity}
%%%%%%%%%%%%%%%%%%%%%%%%%%%%%%%%%%%%%%

If the joint channel $\Cc$ in \eqref{eq:channels} is unitary, then the marginal channels $\Cc_A$ and $\Cc_B$ are called \emph{conjugate channels}, or \emph{complementary channels}. Hence, by definition, conjugate channels are compatible. Often conjugate channels are considered in a situation where $\Cc_A$ is used to transmit a quantum state and its conjugate channel $\Cc_B$ describes what happens on the environment, in particular, how the information (either classical, or quantum) encoded in $\varrho$ is diluted into the environment. 
Therefore, it is not surprising that transmission capacities of complementary channels are closely related \cite{KiMaNaRu07} and possess some common qualitative features, e.g. the additivity of capacities \cite{Holevo05}.

The special role of conjugate pairs of channels compared to other compatible pairs derives from the fact that a unitary channel cannot loose any information on the input state. Therefore, if $\Cc_A$ destroys some information, there is a corresponding flow of information to the environment. 
A neat quantitative formulation of this information-disturbance trade-off is the following \cite{KrScWe08ieee}:
\begin{equation}\label{eq:KSW}
\frac{1}{4} \inf_\mathcal{D} \no{ \Cc_A \circ \mathcal{D} - id }_{cb}^2 \leq   \no{ \Cc_B  - \mathcal{A}_\xi }_{cb} \leq 2 \inf_\mathcal{D} \no{ \Cc_A \circ \mathcal{D} - id }_{cb}^{\frac{1}{2}} \, ,  
\end{equation}
where all channels are written in the Heisenberg picture, the infimum is taken over all decoding channels $\mathcal{D}$ and $\mathcal{A}_\xi$ is a completely depolarizing channel for some fixed state $\xi$, i.e., $\mathcal{A}_\xi(\varrho)=\xi$ for all $\varrho$.
This result shows that if we can find a decoding channel $\mathcal{D}$ such that almost all the information can be retrieved from the output of $\Cc_A$, then the conjugate channel $\Cc_B$ is well approximated
by a completely depolarizing channel, hence the information flow to the environment is small.

If $\Cc_A$ is a unitary channel, then it is reversible and can be perfectly decoded. 
From \eqref{eq:KSW} we conclude that then the conjugate channel $\Cc_B$ must be a completely depolarizing channel. And \emph{vice versa}, if we start from the assumption that $\Cc_B$ is a completely depolarizing channel, then $\Cc_A$ must be a perfectly decodable.
As a consequence, two completely depolarizing channels are not conjugate.
However, as they clearly are compatible (as they are trivial devices), we have demonstrated that \emph{there are compatible channels that are not conjugate}.

Let us remark that, perhaps surprisingly, a channel can be conjugated 
with itself. To see this, let $\Cc$ be a channel that is induced by a controlled unitary operator $U_{\rm ctrl}$ acting on two qubits,
\begin{equation}
\Cc(\varrho\otimes\varrho_0)=
U_{\rm ctrl}(\varrho\otimes\varrho_0)U_{\rm ctrl}^\dagger
\end{equation}
 with
\begin{equation}
U_{\rm ctrl}=I\otimes\ket{\varphi}\bra{\varphi}+\sigma_z\otimes\ket{\varphi_\perp}\bra{\varphi_\perp} \, ,
\end{equation}
where $\{\varphi,\varphi_\perp\}$ is an orthonormal basis. 
Setting $\varrho_0=\ket{\varphi_+}\bra{\varphi_+}$ with
$\varphi_{\pm}=(\varphi\pm\varphi_\perp)/\sqrt{2}$ we obtain
\begin{align}
\Cc_A(\varrho)&=\frac{1}{2}\varrho+\frac{1}{2}\sigma_z\varrho\sigma_z=
{\rm diag}_z(\varrho)\\
\Cc_B(\varrho)&=\frac{1}{2}[I+\tr{\varrho\sigma_z}
(\ket{\varphi_+}\bra{\varphi_+}-\ket{\varphi_-}\bra{\varphi_-})].
\end{align}
If we further choose $\ket{\varphi}=\ket{+}$ and $\ket{\varphi_\perp}=\ket{-}$, 
then $\Cc_A=\Cc_B={\rm diag}_z$. 

This example can be generalized to arbitrary dimension. In particular, consider
a set of $d$ mutually commuting unitary operators $U_1,\dots,U_d$ such that
$\tr{U_j^\dagger U_k}=d\delta_{jk}$. Choose a vector state $\varphi$. Then
the vectors $\ket{\varphi_j}=U_j\ket{\varphi}$ form an orthonormal basis 
$\mathcal{B}$
of the Hilbert space $\mathcal{H}_d$. Further, let us define a control unitary
operator $U_{\rm ctrl}=\sum_j U_j\otimes\ket{\varphi_j}\bra{\varphi_j}$ and
$\varrho_0=\ket{\overline{\varphi}}\bra{\overline{\varphi}}$, where
$\ket{\overline{\varphi}}=(1/\sqrt{d})\sum_j\ket{\varphi_j}$ is equal
superposition of all basis vector states. 
It follows that
\begin{equation}
\Cc_A(\varrho)={\rm tr}_B[U_{\rm ctrl}(\varrho\otimes\varrho_0)U_{\rm ctrl}^\dagger]=
\frac{1}{d}\sum_j U_j\varrho U_j^\dagger\,,
\end{equation}
and
\begin{equation}
\Cc_B(\varrho)={\rm tr}_A[U_{\rm ctrl}(\varrho\otimes\varrho_0)U_{\rm ctrl}^\dagger]=
\frac{1}{d}\sum_{j,k}\tr{U_j\varrho U_k^\dagger}\ket{\varphi_j}\bra{\varphi_k}\,.
\end{equation}
Using the fact that 
\begin{equation}
\tr{U_j\varrho U_k^\dagger}=
\sum_n \bra{\varphi} U_n U_j\varrho U_k^\dagger U_n^\dagger\ket{\varphi}
= \bra{\varphi_j}\left(\sum_n U_n\varrho U_n^\dagger\right)\ket{\varphi_k}
\end{equation}
we obtain
\begin{eqnarray*}
\Cc_B(\varrho)&=&\frac{1}{d}\sum_{j,k} \ket{\varphi_j}\bra{\varphi_j}
\left(\sum_n U_n\varrho U_n^\dagger\right)\ket{\varphi_k}\bra{\varphi_k}
\\ &=&\frac{1}{d}\sum_n U_n\varrho U_n^\dagger = \Cc_A (\varrho)
= {\rm diag}_{\mathcal{B}}(\varrho)\,.
\end{eqnarray*}

In conclusion, every diagonalizing channel is conjugate with itself. 
Combining this fact with the last paragraph of the previous subsection 
we conclude that the \emph{diagonalizing channels ${\rm diag}_{\mathcal{B}}$ 
and ${\rm diag}_{\mathcal{B^\prime}}$ are  compatible if and only if $\mathcal{B} = \mathcal{B^\prime}$.}

%%%%%%%%%%%%%%%%%%%%%%%%%%%%%%%%%%%%%%
\subsection{Incompatibility of quantum observable and channel}
%%%%%%%%%%%%%%%%%%%%%%%%%%%%%%%%%%%%%%

A \emph{quantum instrument} represents the mathematical tool enabling us to go beyond purely statistical description of quantum measurements \cite{OQP97}, \cite{MLQT12}. 
It includes not only the probabilities for measurement outcomes but also the effect of the measurement process on the state of the measured object
conditioned on the recorded outcome. 
The most general state transformation
is described by a quantum operation, which is a completely positive 
trace-non-increasing linear map acting on the set of trace class 
operators. 
Quantum instruments are then normalized operation-valued measures. 
In particular, a measurement with finite number of outcomes is described 
by mapping $x\mapsto\Ii_x$, where for each outcome $x\in\Omega$ 
the transformation $\Ii_x$ is a quantum operation. The probability of outcome $x$ given
the initial state $\varrho$ is $p_\varrho(x)=\tr{\Ii_x(\varrho)}$ and the
conditioned state equals 
$\varrho_x=\frac{1}{p_\varrho(x)}\Ii_x(\varrho)$. 
Clearly, the condition $\sum_x p_\varrho(x)=1$ is guaranteed if $\Ii_\Omega$ 
is trace-preserving. 
In conclusion, quantum instrument can be understood as a device
having a state $\varrho$ at its input and producing two outputs: 
i) probability distribution (described by some observable) and 
ii) an average output state (described by a channel). 
Every quantum instrument has a representation in the form of a measurement process \cite{Ozawa84}, and this fact justifies their use in the description of quantum measurements. 

A quantum instrument is, by definition, a device with one input port and two output ports, one classical and one quantum.
Hence, the connection to the general definition in Subsec. \ref{sec:definition} can be directly applied, and we thus conclude that an observable $\Mo$ and a channel $\Cc$ are compatible exactly if there exist a quantum instrument $\Ii$ such that $\tr{\varrho\Mo(x)}=\tr{\Ii_x(\varrho)}$ and $\Cc(\varrho)=\Ii_\Omega(\varrho)$ for all outcomes $x$ and input states $\varrho$.
This type of compatibility was already investigated in \cite{Ozawa84}, \cite{Ozawa85}.  
The concept of an instrument gives rise also to another sort of compatibility, as two operations may or may not belong to the range of a single instrument \cite{HeReStZi09}.
Further, the limitations on approximate joint measurability of two measurements become different if we our aim is to approximate not only measurement outcome probabilities but also state transformations \cite{HeJiReZi10}.

To demonstrate the compatibility relations between observables and  channels and the mathematical form of instruments, we write explicitly the joint devices for an observable and a channel when one of them is a trivial devices.
First, any channel is compatible with any trivial observable, meaning that whatever we do with the input system, we can additionally toss a coin.
An instrument for a channel $\Cc$ and a trivial observable $\To$ is simply
\begin{equation}\label{eq:instrument-trivial}
\Ii_x(\varrho)= \Cc(\varrho) \To(x) \, .
\end{equation}
Second, any observable is compatible with any trivial channel, meaning that any observable can be measured in a totally destructive way.
An instrument for an observable $\Mo$ and a completely depolarizing channel $\Ac_\xi$ is
\begin{equation}\label{eq:instrument-with-dc}
\Ii_x(\varrho)= \tr{\varrho \Mo(x)} \xi \, .
\end{equation}
A general instrument has, of course, more complicated structure and it need not be decomposable as the previous instruments.
For some examples of instrument arising from realistic measurement models we refer to \cite{OQP97}.  

There is a slightly more general class of channels that can be written for any observable. 
Let $\Mo$ be an observable and fix a state $\xi_x$ for each possible outcome $x$.
We can then define an instrument as
\begin{equation}
\Ii_x(\varrho)= \tr{\varrho \Mo(x)} \xi_x \, .
\end{equation}
This instrument describes a measurement process where we measure  $\Mo$ and, depending on the obtained outcome, prepare one of the states $\xi_x$.
The channel $\Ii_\Omega$ deriving from this instrument maps an input state $\varrho$ into the convex mixture $\sum_x p_\varrho(x) \xi_x$.
Typically an observable has a variety of other kind of compatible channels as well, but \emph{if each operator $\Mo(x)$ is rank-1, then $\Mo$ has no other kind of compatible channels} \cite{HeWo10}.
The destructive nature of measurements of rank-1 observables is connected to the partial order of observables where rank-1 observables are maximal; this aspect will be explained in Sec. \ref{sec:order}.

The mathematical formalism of instruments allows one to formulate the notion of sequential measurements \cite{DaLe70},\cite{BuCaLa90}.
A sequential measurement of two observables $\Ao$ and $\Bo$ gives rise to a joint measurement of $\Ao$ and a deformed version $\Bo'$ of $\Bo$.
Clearly, a sequential measurement is a special kind of realization of a joint measurement, so the implemented observables $\Ao$ and $\Bo'$ must always be compatible. 
However, if  $\Ao$ and $\Bo$ are compatible, then one can try to foresee the disturbance caused by the first measurement and measure some other observable $\Co$ instead of $\Bo$ is order to implement a joint measurement of $\Ao$ and $\Bo$.
It was shown in \cite{HeMi15} that there is, in fact, a fixed instrument for $\Ao$ such that all observables compatible with $\Ao$ can be obtained by measuring sequentially some observable after $\Ao$. 
Hence, it is possible to perform a measurement of a quantum observable in a way that does not
disturb the subsequent measurements more than is dictated by joint measurability.

%%%%%%%%%%%%%%%%%%%%%%%%%%%%%%%%%%%%%%
\subsection{No information without disturbance}\label{sec:no-info}
%%%%%%%%%%%%%%%%%%%%%%%%%%%%%%%%%%%%%%

It is one of the main features of quantum theory that the disturbance
caused by a measurement must be irreversible if the measurement provides some nontrivial information about the system. This statement is known as \emph{no information without disturbance}. 
In the language of incompatibility this means that \emph{a unitary channel is incompatible with any nontrivial observable}. 

A proof of this statement can be found in many textbooks 
(see e.g. \cite{MLQT12}). Here we sketch a simple argument.  
Suppose that a unitary channel 
$\varrho \mapsto U \varrho U^*$ is compatible with an observable $\Ao$. 
This means that there is an instrument $\Ii$ such that 
$\Ii_x^*(\id)=\Ao(x)$ and $\sum_x \Ii_x(\varrho) = U\varrho U^*$ for all $x$ and $\varrho$. 
Let then $\Bo$ be any observable.
By measuring the observable $U\Bo U^*$ afterwards, 
we can realize a simultaneous measurement of $\Ao$ and $\Bo$.
As $\Bo$ is arbitrary, we conclude that each operator $\Ao(x)$ commutes with all projections (recall Subsec. \ref{sec:comma}).
Hence, $\Ao$ must be a trivial observable.

Although universally \emph{no information} is a necessary requirement 
for \emph{no disturbance} there are cases when some information can be extracted
without causing disturbance. For example, this may happen when quantum 
systems are used to encode classical information, hence, orthogonal vector 
states $\varphi_0, \varphi_1$ are selected to represent one bit of information.  
Denote by $P_j = \kb{\varphi_j}{\varphi_j}$ the associated one-dimensional projections. These projections describe pure states, but they also determine an observable $\Mo(j)=P_j$. Since $\tr{P_j\Mo(k)}=\delta_{jk}$, we see that  $\Mo$ can perfectly discriminate the states $\varphi_0,\varphi_1$ from just one measurement outcome. Further, if we choose a measurement of $\Mo$ implementing the L\"uders instrument $\Ii^L$ given as
\begin{align}\label{eq:luders}
\Ii^L_j(\varrho) = \Mo(j)\varrho \Mo(j) \, , 
\end{align}
then we have $\Ii^L_\Omega(P_j) = P_j$, implying that the 
states $P_j$ are not disturbed. In conclusion, \emph{if the set of input 
states is restricted, then it is possible to retrieve information 
without disturbance}. 
In the area of quantum measurement theory, this is related to the possibility of measurements of the first kind \cite{BuCaLa90}.

The previous simple example resemblances classical setting as all operators commute, and for this reason the conclusion is not that surprising although good to bear in mind. However, at the same time (without the apriori information on states) the measurement in \eqref{eq:luders} is highly state disturbing if its action is considered on different collection of states. In particular, let us consider two pairs of orthogonal states $Z:= \{ \varphi_0, \varphi_1\}$ and 
$X:= \{ \psi_0, \psi_1\}:= \{\frac{1}{\sqrt{2}}(\varphi_0 + \varphi_1), 
\frac{1}{\sqrt{2}}(\varphi_0 - \varphi_1)\}$. 
While L\"uders measurement of $\Mo$ extracts perfectly   
the bit of information encoded in $Z$ without destroying the states, 
it completely spoils the bit of information encoded in states 
$X$. 
In fact we have  $\Ii^L_{\Omega} ( |\psi_j \rangle \langle \psi_j|) 
= \frac{1}{2}\id$ for $j=1,2$.
There are numerous quantitative trade-off relations for the noise and disturbance in the measurement of two quantum observables. 
For some interesting recent developments that have an incompatibility twist, we refer to \cite{BuHaOzWi14}, \cite{MaSr14a}, \cite{ReSc14}. 

An important application of ``no information without disturbance'' is in the area of quantum security, where it is used to guarantee the security of \emph{quantum key distribution} (QKD) protocols. 
Let us briefly recall one of the most simplest QKD protocols known as B92, first described in \cite{Bennett92}. 
B92, as any QKD protocol, has three phases: 
\begin{enumerate}
\item[(i)] establish perfectly correlated strings (representing the raw keys) 
between Alice and Bob, 
\item[(ii)] verify the presence of the eavesdropper (comparing part of the keys),
\item[(iii)] purify the key (if possible).
\end{enumerate}
The first phase is employing the
so-called unambiguous state discrimination (USD) procedure that either 
reliably identifies non-orthogonal states, or results in an inconclusive 
outcome \cite{Chefles00}. 
So Alice prepares randomly
one of the non-orthogonal vector states $\psi$ or $\phi$ (representing bit values 
0, or 1, respectively), and sends it to Bob. 
Bob performs USD measurement and publicly announce when inconclusive outcome was recorded. 
Both Alice and Bob remove these bits from their strings to
obtain a perfectly correlated string of bits each. 
Quantum incompatibility is relevant in the second phase. 
In this phase Alice and Bob compare random bits from their raw keys. 
Observation of any error implies that the protocol was not implemented 
perfectly and no one can say whether the observed imperfections are due 
to some eavesdropper, or of some other (less dangerous) origin. The goal
of the eavesdropper is to learn the key while being undetected. However,
the act of learning is necessarily related to a measurement process, whereas
the detection is impossible only when no disturbance occurs during the
eavesdropping process. Therefore, ``no information without disturbance'' 
feature implies that any curious eavesdropper will be detected. 
Naturally, in practice the situation is never ideal and the protocol should tolerate some degree of disturbance. 
For practical QKD it is important to understand which actions are tolerable in a sense that the third phase could correct the disturbances without compromising the security \cite{Renner08}.

%%%%%%%%%%%%%%%%%%%%%%%
\subsection{Incompatible process positive operator valued measures}
\label{sec:ppovm}
%%%%%%%%%%%%%%%%%%%%%%%

In this section we will illustrate the concept incompatibility in a 
rather nonstandard framework. In particular, we will introduce the 
compatibility questions for measurements on quantum channels. Let us stress that
the incompatibility of processes is not yet explored in much details,
therefore, in this section we will only illustrate several simple features 
that makes the incompatibility of processes qualitatively different from 
the typically considered state-based incompatibility. Namely, we will see 
that for processes the commutativity is not sufficient to guarantee 
the compatibility. Moreover, the commutative pair of process observables 
is among the most incompatible pair of observables and the theoretical
maximum (discussed in Subsec.~\ref{sec:degree}) is achieved 
for quantum system of arbitrary dimension.

\begin{figure}\begin{center}
\includegraphics[width=6cm]{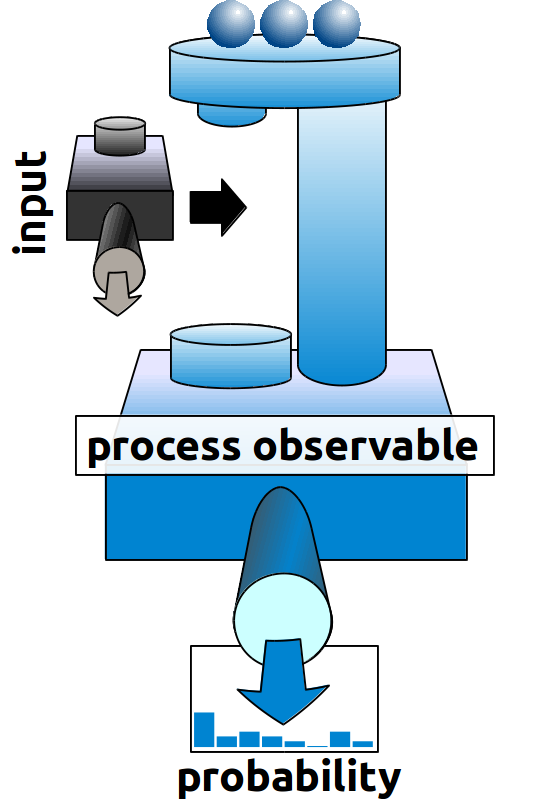}
\caption{Process measurement device (process observable) accepts
processes at its input and produces a probability distribution
of observed events at its output. These events are composed of
a preparation of the probe system and its measurement.}
\end{center}
\end{figure}

Let us recall that quantum channels (for finite dimensional systems) can be 
represented by Choi-Jamiolkowski operators \cite{MLQT12}, i.e. 
$$
\mathcal{S}_{\rm chan}\equiv\{\omega\in\mathcal{S}(\hi\otimes\hi): d{\rm tr}_2[\omega]=I\} \, .
$$
By definition a \emph{process effect} is any affine mapping 
$e:\mathcal{S}_{\rm chan}\to[0,1]$. It turns out \cite{Jencova12} 
they can be associated with positive operators $M\in \mathcal{L}(\hi\otimes\hi)$ satisfying the relation
$$
O\leq M\leq \xi\otimes I\,,
$$
for some density operator $\xi\in\mathcal{S}(\hi)$. Consequently, the
observables on channels are represented by sets of process effects 
$M_1,\dots,M_n$ (forming
a process POVM \cite{Ziman08a}, or 1-tester \cite{ChDaPe09pra}) 
with the normalization $\sum_j M_j=\xi\otimes I$.
All effects in the range of process POVM are bounded by the same density operator 
$\xi$, i.e., $M_j\leq\xi\otimes I$. 
The probability of observing 
the experimental event described by process effect $M_j$ providing that 
channel $\omega\in\mathcal{S}_{chan}$ is tested is given by the Born-like 
formula $p_j(\omega)=\tr{\omega M_j}$.

Consider a pair of two-outcome process observables 
\begin{eqnarray}
\nonumber 
\Dev_\Mo&:& M_0=P_0\otimes P_0,\quad M_1=P_0\otimes P_1 \\
\Dev_\No&:& N_0=P_1\otimes P_0,\quad N_1=P_1\otimes P_1\,,
\end{eqnarray} 
where $P_0=\ket{0}\bra{0}$
and $P_1=\ket{1}\bra{1}$. By definition these process 
observables are incompatible if there exist a process observable 
$\Dev_G$ with outcomes $G_{jl}$ such that $G_{j0}+G_{j1}=M_j$ and 
$G_{0l}+G_{1l}=N_l$. The normalization of $M_j$ and $N_l$
implies that $G_{jl}\leq P_0\otimes I$ and $G_{jl}\leq P_1\otimes I$,
hence, $G_{jl}=P_0\otimes g_{jl}=P_1\otimes g_{jl}^\prime$ for some
positive operators $g_{jl}$ and $g_{jl}^\prime$. However, this is 
possible only if $g_{jl}=g_{jl}^\prime=O$ (implies $G_{jl}=O$) for 
all $j,l$. In conclusion, the process observables $\Dev_\Mo$ and $\Dev_\No$
are incompatible although they are commuting, i.e. $[\Dev_\Mo,\Dev_\No]=0$
(meaning all process effects are commuting).
In other words, for measurements on quantum processes we are coming
with the following rather unexpected conclusion: the commutativity 
does not imply the compatibility. This suggests that questions on 
compatibility of measurements on processes are not reducible to 
analogous questions for measurements of states.

The process observable $\Dev_\Mo$ represents an experiment in which the qubit 
process is applied on the initial state $\ket{0}$ and the output 
is measured by projection-valued observable $\sigma_z$. Similarly,
$\Dev_\No$ describes almost the same experiment, only the initial state 
is chosen to be $\ket{1}$. The incompatibility has relatively clear 
intuitive meaning. It simply says that process cannot be probed
simultaneously by two orthogonal pure states although formally 
they are commuting. Let us stress that this is true for any pair
of pure states, however, for non-orthogonal pair the resulting 
devices are non-commuting.

Moreover, as it is shown in \cite{SeReChZi15} making these process
observables compatible requires maximal possible addition of noise. 
Let us denote by $\Dev_\mathbf{I}$ the trivial process observable (compatible
with any other process observable). Then
the observables $\Dev_{\Mo,q}=q\Dev_\Mo+(1-q)\Dev_\mathbf{I}$
and $\Dev_{\No,q}=q\Dev_\No+(1-q)\Dev_\mathbf{I}$ are compatible only 
if $q\leq 1/2$, which is the worst case compatibility. Any pair of process
observables is compatible at this fraction of added noise. However, 
let us recall from Sec.~\ref{sec:degree} that for the usual observables the
trivial value $\half$ cannot be improved only in the case of some special pairs 
of observables in infinite-dimensional Hilbert space, 
whereas for process observables already the two-dimensional case is sufficient to host the maximally incompatible process measurement devices.

%%%%%%%%%%%%%%%%%%%%%%%%%%%%%%%%%%%
\section{Order theoretic characterization of quantum incompatibility}\label{sec:order}
%%%%%%%%%%%%%%%%%%%%%%%%%%%%%%%%%%%

%%%%%%%%%%%%%%%%%%%%%%%%%%%%%%%%%%%
\subsection{Preordering of devices}\label{sec:prelimorder}
%%%%%%%%%%%%%%%%%%%%%%%%%%%%%%%%%%%

As in Subsec. \ref{sec:prelim}, we consider devices with a fixed input space $\mathcal{S}$ but arbitrary output space. For two devices $\Dev_1: \mathcal{S} \to \mathcal{S}_1$ and $\Dev_2: \mathcal{S} \to \mathcal{S}_2$, 
we write $\Dev_1 \pgeq \Dev_2$ if there exists a third device $\Dev_{12}: \mathcal{S}_1 \to \mathcal{S}_2$ 
such that
\begin{align}
\Dev_{12} \circ \Dev_1 = \Dev_2 \, ,
\end{align}
where $\circ$ denotes function composition of two mappings.
The physical meaning of $\Dev_1 \pgeq \Dev_2$ is that $\Dev_2$ can be simulated by using $\Dev_1$ and $\Dev_{12}$ sequentially.

It follows from the definition that if $\Dev_1 \pgeq \Dev_2$ and $\Dev_2 \pgeq \Dev_3$, then $\Dev_1 \pgeq \Dev_3$.
Moreover, $\Dev_1 \pgeq \Dev_1$ since the identity map is a possible device. 
We conclude that on any subset of devices, the relation $\pgeq$ is a preorder.
In typical considerations there are devices that can simulate each other without being the same (e.g. reversible channels), so the preorder $\pgeq$ fails to be a partial order.
We now obtain a direct consequence of the definition of compatibility:
\begin{quote}
\emph{If $\Dev_1$ and $\Dev_2$ are two compatible devices and some other devices $\Dev'_1$ and $\Dev'_2$ satisfy
$\Dev_1 \pgeq \Dev'_1$ and $\Dev_2
\pgeq \Dev'_2$, then $\Dev'_1$ and $\Dev'_2$  are compatible.}
\end{quote}
Namely, let $\Dev$ be a joint device of $\Dev_1$ and $\Dev_2$.
Let $\Dev'_1$ and $\Dev'_2$ be such that
$\Dev'_1 = \Dev_{11} \circ \Dev_1$ and $\Dev'_2 = \Dev_{22} \circ \Dev_2$ for some devices $\Dev_{11}$ and $\Dev_{22}$.
We set
\begin{equation}
\Dev' = (\Dev_{11} \otimes \Dev_{22}) \circ \Dev \, . 
\end{equation}
Then $\Dev'$ is a joint device of $\Dev'_1$ and $\Dev'_2$.

We denote $\Dev_1\simeq \Dev_1'$ if both $\Dev_1 \pgeq \Dev_1'$ and $\Dev_1'\pgeq \Dev_1$ hold. 
From the previous observation we conclude the following:
\begin{quote}
\emph{Let $\Dev_1$ and $\Dev_1'$ be two devices and $\Dev_1 \simeq \Dev'_1$.
A device $\Dev_2$ is compatible with $\Dev_1$ if and only if $\Dev_2$ is compatible with $\Dev'_1$.}
\end{quote}
This observation indicates that a more natural setting for the incompatibility relation are equivalence classes of devices rather than single devices.
In the following subsections we demonstrate how naturally many incompatibility results on quantum devices follow from this order structure.

%%%%%%%%%%%%%%%%%%%%%%%%%%
\subsection{Preordering of quantum observables}
%%%%%%%%%%%%%%%%%%%%%%%%%%%

 In order to formulate the preorder between observables in the sense of Subsec. \ref{sec:prelimorder}
we have to first recall the notion of a \emph{classical channel}, which is an affine map from $\mathcal{P}(\Omega)$ to $\mathcal{P}(\Omega')$, where $\mathcal{P}(\Omega)$ denotes
the set of all probability distributions on a set $\Omega$. 
A classical channel be conveniently presented by a real valued function $(x,y) \mapsto p(x \mid y)$ on the Cartesian product $\Omega \times \Omega'$ satisfying $p(x \mid y) \geq 0$ and $\sum_x p(x \mid y) = 1$.
A probability distribution $\nu$ on $\Omega$ is then mapped into a probability distribution $\nu'$ on $\Omega'$,
\begin{align}
\nu'(y) = \sum_x p(y \mid x) \nu(x) \, .
\end{align}
Given two observables $\Mo$ and $\No$, we denote $\Mo\pgeq\No$ if there exists a classical channel $p$ such that
\begin{equation}\label{eq:smearing}
\No(y)=\sum_x p(y \mid x) \Mo(x) 
\end{equation}
for all $y\in\Omega_{\No}$.
Our formulation of incompatibility in Subsec. \ref{sec:functional} can be now restated as follows:
\begin{quote}
\emph{
Observables $\Mo_1,\ldots,\Mo_n$ are compatible if and only if 
there exists an observable $\Mo$ such that $\Mo_j\pleq \Mo$ for every $j=1,\ldots,n$.}
\end{quote}

As in Subsec. \ref{sec:prelimorder}, we denote $\Mo\simeq \No$ if both $\Mo \pgeq \No$ and $\No\pgeq \Mo$ hold. 
Then $\simeq$ is an equivalence relation and the equivalence class of $\Mo$ is denoted by $[\Mo]$. 
We introduce the set of equivalence classes $\obs^\sim := \obs / \simeq$ and the preorder $\pleq$ then induces a partial order
 $\pleq$ on $\obs^\sim$  by $[\Mo] \pleq [\No]$ if and only if $\Mo \pleq \No$. 
(We use the same symbol $\pleq$ for these two different relations, but this should not cause a confusion.)
From the fact that incompatibility properties are the same for two observables belonging to a same equivalence class we can already conclude some useful facts.
In particular, the commutativity results stated in Subsec. \ref{sec:comma} for sharp observables remain unchanged for an observable that is equivalent to a sharp observable. 
A sharp observable can be a smearing of another observable only if the corresponding classical channel takes only the values $0$ and $1$ \cite{JePu07}, so a typical example of an observable $\Mo$ that is equivalent with a sharp observable $\Po$ is of the form
\begin{equation}
\Mo(x,y)= p(x,y) \Po(x) \, , 
\end{equation}
where $0 \leq p(x,y) \leq 1$.

As shown in \cite{MaMu90a}, \emph{the least element in $\obs^\sim$ consists of all trivial observables}, while \emph{there is no greatest element in $\obs^\sim$}. 
The first statement can be equivalently formulated as: \emph{a quantum observables that is compatible with all other observables is trivial}, while the second statement can be equivalently formulated as: \emph{not all quantum observables are compatible}. 
The order theoretic structure of $\obs^\sim$ is thus directly reflected in the compatibility relation of observables.

While there is no greatest element in $\obs^\sim$ , there are maximal elements, i.e., elements that are not below any other element.
An individual observable $\Mo$ is called maximal if it belongs to a maximal equivalence class. 
Hence, $\Mo$ is maximal if and only if $\Mo \pleq \No$ implies $\Mo\simeq \No$.
In the case of a finite dimensional Hilbert space, maximal observables are exactly those whose all nonzero operators are rank-1 \cite{MaMu90a}.
We thus conclude that \emph{two rank-1 observables are compatible if and only if they are equivalent.}

%%%%%%%%%%%%%%%%%%%%%%%%%%
\subsection{Preordering of quantum channels}
%%%%%%%%%%%%%%%%%%%%%%%%%%%
 
In the following it is more convenient to use the Heisenberg picture for quantum channels.
A quantum channel is then defined as a normal completely positive map $\Cc: \lk \to \lh$ satisfying $\Cc(\id_{\hik})=\id_{\hi}$, where $\hik$ is the output Hilbert space. 
The Schr\"odinger picture description $\Cc^S$ of a channel $\Cc$ can be obtained from the relation
\begin{equation}
\tr{\Cc^S(\varrho)T}=\tr{\varrho \Cc(T)} \, , 
\end{equation}
required for all states $\varrho\in\lh$ and operators $T\in\lk$.

We denote by $\chan$ the set of all channels from an arbitrary output space $\lk$ to the fixed input space $\lh$.
The sequential implementation of channels is opposite in the Heisenberg picture as in the Schr\"odinger picture, so for two channels $\Cc_1, \Cc_2\in \chan$ we have $\Cc_1 \pleq \Cc_2$ if there exists a channel $\mathcal{E}$ such that $\Cc_1 =\Cc_2 \circ \mathcal{E}$. 
As explained earlier, it is often convenient to work on the level of equivalence classes of channels, and we denote
$\chan^\sim:=\chan / \sim$. 
In the partially order set $\chan^\sim$, there exists the least element and the greatest element. 
The least equivalence class consist of all complete depolarizing channels, which in the 
the Heisenberg picture are
\begin{equation}
\Lambda_{\eta}(T)=\tr{\eta T}\id \, ,
\end{equation} 
where $\eta$ is some fixed state.
The greatest element in $\chan^\sim$ is the equivalence class of the identity channel $id$. 

The compatibility of two channels can be neatly expressed in terms of conjugate channels.
In this context a conjugate channel $\Cc^c$ of a channel $\Cc$ from $\mathcal{L}(\hik')$ to $\mathcal{L}(\hik)$ is 
introduced by using Stinespring representation $(V, \hik')$ of $\Cc$, and then 
\begin{eqnarray*}
\Cc(A) = V^*(A \otimes \id)V \, , \quad \Cc^c(B) = V^*(\id \otimes B)V \, . 
\end{eqnarray*} 
Any channel has a unique equivalence class of its conjugate channel irrespective of the choice of Stinespring representation \cite{HeMi15}. 
Therefore, we can understand the conjugation as a function $[\Cc] \mapsto [\Cc]^c$ on $\chan^\sim$.
We then have the following characterization of compatibility:
\begin{quote}
\emph{Two channels $\Cc_1$ and $\Cc_2$ are compatible if and only if 
$[\Cc_2] \pleq [\Cc_1]^c$ (or equivalently $[\Cc_1]
\pleq [\Cc_2]^c$) holds.} 
\end{quote}

While the 'if' part in this statement is trivial, the 'only if' may require some explanation. 
Suppose that $\Cc_1$ and $\Cc_2$ are compatible. 
Then there exists a channel $\Cc$ such that 
$\Cc(A\otimes \id) = \Cc_1(A)$ and 
$\Cc(\id \otimes B) = \Cc_2(B)$ hold. 
Let us denote Stinespring representation of $\Cc$ by 
$(V, \hik)$, so that
\begin{equation}
\Cc(A\otimes B) = V^*(A\otimes B\otimes \id_{\hik})V \, .
\end{equation}
Now $\Cc_1^c$ is written as 
$(\Cc_1^c)(B \otimes C) = V^*(\id \otimes B\otimes C)V$. 
We define a channel $\Ec$ by 
$\Ec(B ) = B\otimes \id_{\hik}$. Then 
$\Cc_1^c \circ \Ec = \Cc_2$ holds.   
This result verifies the intuition that for a given channel $\Cc_1$, its conjugate channels are the best channels still compatible with $\Cc_1$.
 
%%%%%%%%%%%%%%%%%%%%%%%%%%%%%%%
\subsection{Incompatibility between observable and channel}
%%%%%%%%%%%%%%%%%%%%%%%%%%%%%%%

Let us denote by $\ch{\Mo}$ the set of all channels compatible with an observable $\Mo$. 
We call a channel compatible with $\Mo$ a $\Mo$-channel. 
It was proved in \cite{HeMi13} that
\emph{there exists a channel $\Lambda_\Mo$ such that  
the set $\chan_{\Mo}$ of all channels compatible with $\Mo$ consists of all channels that are below $\Lambda_\Mo$, i.e.,
\begin{equation}
\chan_{\Mo}=\{ \Cc \in \chan \, |\,  \Cc \pleq \Lambda_\Mo \} \, .
\end{equation}}
From the physical point of view, this result tells that there is a specific channel $\Lambda_\Mo$ among all $\Mo$-channels, and all other $\Mo$-channels can be obtained from $\Lambda_\Mo$ by applying a suitable channel after the measurement. 
It is even justified to call $\Lambda_\Mo$ a \emph{least disturbing} $\Mo$-channel since an additional channel after it cannot decrease the caused disturbance.

The mathematical form of  $\Lambda_\Mo$ is simple to write by using the Naimark dilation of $\Mo$. 
Namely, let $(\hik,\hat{\Mo},V)$ be a Naimark dilation of $\Mo$, i.e., $\hik$ is a Hilbert space, $V$ is an isometry 
$V: \hi \to \hik$, and $\hat{\Mo}$ is a sharp observable  
on $\hik$ satisfying 
\begin{equation}
V^*\hat{\Mo}(x)V = \Mo(x) \, .
\end{equation}
Then
\begin{equation}
\Lambda_\Mo(\varrho) = \sum_x \hat{\Mo}(x) V \varrho V^* \hat{\Mo}(x)
\end{equation}
for all input states $\varrho$.
Of course, any channel equivalent with $\Lambda_\Mo$ has the same order property, so the least disturbing channel is unique only up to an equivalence class. 

As one would expect, more noise on the observable $\Mo$ allows less disturbance in its least disturbing channel $\Lambda_\Mo$.
In fact, it was shown in \cite{HeMi13} that the following conditions are equivalent:
\begin{itemize}
\item[(i)] $\Mo\pleq\No$ 
\item[(ii)] $\ch{\No}\subseteq\ch{\Mo}$
\item[(iii)] $\Lambda_\No \pleq \Lambda_\Mo$
\end{itemize}
This result can be seen as a qualitative noise-disturbance relation; even without any quantification of disturbance we can clearly say that the condition $\ch{\No}\subseteq\ch{\Mo}$
 means that $\Mo$ allows less or equally disturbing measurement than the least disturbing measurement of $\No$.

%%%%%%%%%%%%%%%%%%%%%%%%%%5
\section{Outlook}\label{sec:outlook}
%%%%%%%%%%%%%%%%%%%%%%%%%%%5

In the early days of quantum theory its founding fathers
realized that measurement statistics of conjugated physical quantities, such as position and momentum, have mutual limitations. 
Formulating the notions of uncertainty principle and complementarity (in a relatively vague form) they discovered the first signs of puzzling phenomenon of quantum incompatibility. 
After hundred years of development in quantum physics and quantum 
technologies the quantum incompatibility has changed its status from a peculiar limitation to the very quantum paradigm. 
In the earlier sections we hope to have demonstrated that the quantum incompatibility is present across many parts of quantum theory and it provides a conceptual way to separate quantum and classical features. 

Although nowadays many aspects of quantum incompatibility are 
understood in details and related applications are being in focus of 
ongoing research programs, there are still parts of quantum theory, where the role of incompatibility is not yet fully recognized. For example, the area 
of incompatibility of measurements of quantum processes is unexplored research territory. 
And because of its qualitative and quantitative 
differences (illustrated in Section \ref{sec:ppovm}) it could have potential 
impact on future quantum applications.
  
In our presentation the phenomenon of incompatibility 
is purely a consequence of the used mathematical framework. We have 
discussed its elementary mathematical properties and 
impacts on physics of quantum systems and related information 
processing. However, understanding conceptually the physical, 
or informational origins of incompatibility is something that 
definitely deserves future attention. The incompatibility as discussed 
in this paper can be studied in any probabilistic 
(toy) theory. 
There seems to be no satisfactory explanation why quantum theory is as incompatible as it is, and a general framework is needed to investigate this question.

%%%%%%%%%%%
\section*{Acknowledgments}
%%%%%%%%%%%

We thank Daniel Reitzner for pointing out that the inequality in \cite{CaHeTo12} can be presented in the symmetric form as in \eqref{eq:lambda}. We are grateful to Tom Bullock and Jussi Schultz for their comments on an earlier version of this paper. 
TM acknowledges JSPS KAKENHI (grant no. 15K04998).
MZ acknowledges the support of projects QUICOST VEGA 2/0125/13 and QIMABOS APVV-0808-12.

%%%%%%%%%%%
\section*{References}
%%%%%%%%%%%

%%%%%%%%%
%%%%%%%%%
\end{document}